\documentclass[prl,twocolumn,superscriptaddress]{revtex4}
\pdfoutput=1
\usepackage{graphicx}
\usepackage{amsmath,graphicx}
\usepackage{epsfig}
\usepackage{paralist}
\usepackage{amsmath,amssymb,mathrsfs}
\usepackage{hyperref}
\usepackage{paralist}
\usepackage{underscore}  
\usepackage{natbib}

\usepackage{color}

\newcommand{\ua}{\uparrow}
\newcommand{\da}{\downarrow}

\def\be{\begin{equation}}
\def\ee{\end{equation}}
\def\bea{\begin{eqnarray}}
\def\eea{\end{eqnarray}}

\newcommand{\tmop}[1]{\ensuremath{\operatorname{#1}}}

\begin{document}

\title{Critical Behaviors of Contact near Phase Transitions}

\author{Y.-Y. Chen}
\affiliation{State Key Laboratory of Magnetic Resonance and Atomic and Molecular Physics, 
Wuhan Institute of Physics and Mathematics, Chinese Academy of Sciences, Wuhan 430071, China}

\author{Y.-Z. Jiang}
\affiliation{State Key Laboratory of Magnetic Resonance and Atomic and Molecular Physics, 
Wuhan Institute of Physics and Mathematics, Chinese Academy of Sciences, Wuhan 430071, China}

\author{X.-W. Guan}
\affiliation{State Key Laboratory of Magnetic Resonance and Atomic and Molecular Physics, 
Wuhan Institute of Physics and Mathematics, Chinese Academy of Sciences, Wuhan 430071, China}
\affiliation{Center for Cold Atom Physics, Chinese Academy of Sciences, Wuhan 430071, China}
\affiliation{Department of Theoretical Physics,
Research School of Physics and Engineering,
Australian National University, Canberra ACT 0200, Australia}

\author{Qi Zhou}
\affiliation{Department of Physics, The Chinese University of Hong Kong, Shatin, New Territories, HK}

\date{\today}



\maketitle

\noindent

{\bf  A central quantity of importance for ultracold atoms is contact, which measures  two-body correlations at short distances in dilute systems. It appears in universal relations among thermodynamic quantities, such as large momentum tails, energy, and dynamic structure factors, through the renowned Tan relations. However, a conceptual question remains open as to whether or not contact can signify phase transitions that are insensitive to short-range physics. Here we show that, near a continuous classical or quantum phase transition, contact exhibits a variety of critical behaviors, including scaling laws and critical exponents that are uniquely determined by the universality class of the phase transition and a constant contact per particle. We also use a prototypical exactly solvable model to demonstrate these critical behaviors in one-dimensional strongly interacting fermions.  Our work establishes an  intrinsic connection between the universality of dilute many-body systems and universal critical phenomena near a phase transition. }

{\bf Introduction}

The notion of contact $\mathcal{C}$ strikingly captures the universality of ultracold atoms. As revealed by the Tan relations \cite{Tan1,Tan2,Tan3} and their expressions in other forms 
\cite{V1,V2,V3,V4},  regardless of the choice of microscopic parameters, a wide range of quantities in dilute systems is governed by $\mathcal{C}$, which characterizes the probability that two particles may be separated by a short distance less than ${\bf d}$. For instance, when ${\bf d}$ approaches zero, the two-body correlation function of two-component fermions in three dimensions follows $\int_{|{\bf x}-{\bf x}'|<d}d{\bf x'}\langle \hat{n}_\uparrow({\bf x})\hat{n}_\downarrow({\bf x}')\rangle\sim  \mathcal{C}|{\bf d}|/(4\pi)$, where  $\hat{n}_{\uparrow}({\bf x})$($\hat{n}_{\downarrow}({\bf x})$) is the density operator at position ${\bf x}$ for spin-up(down) particles.   While the definition of contact is apparently independent of the many-body phase the system is exhibiting, there is much interest in exploring the behaviour of contact near a phase transition \cite{Jin1, Jin2, Jin3, Vale,T1,T2,T3,T4,T5,Braaten}.  The success of such efforts will  significantly deepen our understanding of the connection between short-ranged two-body correlations and phase transitions,  which are generally believed to be disentangled from each other, since the latter is insensitive to the details of short-range physics.

Even though the contact of strongly interacting fermions remains finite in both the normal and the superfluid phase, experimental studies have provided evidence indicating that it gets enhanced near the superfluid transition temperature.  However, due to a lack of resolution, it is unclear whether contact exhibits any critical features near the transition point. On the theoretical side, it is extremely  difficult to exactly calculate the contact of strongly interacting fermions near the transition temperature and certain approximations have to be made. Theoretical approaches based on different techniques lead to contradictory results \cite{Jin3}, ranging from a kink to a discontinuous jump of the contact near the transition point. Therefore, it is of fundamental importance to provide a concrete answer for the relation between contact and phase transitions.  

In this work, our approach is to derive exact results on the behaviour of contact near a classical or quantum phase transition, based on a fundamental thermodynamic relation that is free from any approximations. These results unambiguously show that contact must display critical behaviors near the transitions and that the corresponding critical behaviors are uniquely determined by the universality class of the phase transition. We use a one-dimensional exactly solvable model of strongly interacting fermions exhibiting exotic quantum phase transitions to demonstrate these critical phenomena. Our exact result for contact is obtained from the Bethe ansatz for a one-dimensional Fermi gas that provides a precise understanding of  critical phenomena beyond the Tomonaga-Luttinger liquid physics.   Whereas our general results apply to all dimensions, this one-dimensional example sheds light on the universal features of contact near a phase transition. \\

\noindent
{\bf Results}

{\bf Critical behaviors of contact in three dimensions}  

We consider the fundamental thermodynamic relation, 
\begin{equation}
dP=nd\mu+s dT+MdH-\frac{\rho_\textrm{s}}{2} dw^2+\it{c}d(a_\textrm{3D}^{-1}), \label{3D}
\end{equation}
where $P$ is the pressure, 
$n$, $s$, $M$, $\rho_\textrm{s}$ and $c\equiv \mathcal{C}/\mathcal{V}$ are the densities of the particles, the entropy, the magnetization, the superfluid, and the contact, respectively, $\mathcal{V}$ is the volume of the system, and $\mu$, $T$, $H$ and $a_\textrm{3D}$ are the chemical potential, temperature, magnetic field and scattering length,  respectively. Compared with the usual definition of contact, here the pre-factor $ \hbar^2/(4\pi m)$ (with $m$ denoting the mass) has been absorbed into $c$.  In this relation, $w=v_s-v_n$ is the difference between the velocity of the superfluid and normal components, which can be generated by slowly rotating the atomic cloud so that the critical velocity of the superfluid is not reached anywhere in the trap. Equation (\ref{3D}) has been used to measure thermodynamic quantities such as the pressure, the equation of state and density-density response function\cite{Ho, Sal1, Sal2,Vale2,Martin}. 

Compared with the original Tan relation $\left(\frac{d E}{d (-a_\textrm{3D}^{-1})}\right)_S=\frac{\hbar^2}{4\pi m}\mathcal{C}$, where $E$ and $S$ are the total energy and entropy of the system, respectively, equation (\ref{3D}) has the advantage of allowing one to directly correlate the contact with phase transitions for both classical and quantum phase ones.
First, as $c$ is a partial derivative of the pressure, i.e., $c=({\partial P}/{\partial a_\textrm{3D}^{-1}} )_{\mu, T, H, w}$, as are $n$, $s$ $M$ and $\rho_\textrm{s}$, equation (\ref{3D}) tells one that  contact near the critical point must exhibit critical behaviour determined by the universality class of the phase transition. In particular, the contact should vary continuously 
across a continuous phase transition. For instance, across the superfluid phase transition of strongly interacting fermions in three dimensions, $c$ is continuous. Second, the Maxwell relations derived from equation (\ref{3D}) show that the derivatives of the contact with respect to $\mu, T, H$ and $w$ exhibit critical behaviour. These Maxwell relations can be written as
\begin{eqnarray}
\left(\frac{\partial c}{\partial \mu}\right)_{T, H, w, a_\textrm{3D}^{-1}}&=&\left(\frac{\partial n}{\partial a_\textrm{3D}^{-1}}\right)_{\mu, T, w, H}\label{n},\\
\left(\frac{\partial c}{\partial T}\right)_{\mu,  H, w, a_\textrm{3D}^{-1}}&=&\left(\frac{\partial s}{\partial a_\textrm{3D}^{-1}}\right)_{\mu, T, w, H }\label{s},\\
\left(\frac{\partial c}{\partial H}\right)_{\mu, T, w, a_\textrm{3D}^{-1}}&=&\left(\frac{\partial M}{\partial a_\textrm{3D}^{-1}}\right)_{\mu, T, w, H}\label{m},\\
\left(\frac{\partial c}{\partial w^2}\right)_{\mu, T, H, a_\textrm{3D}^{-1}}&=&-\frac{1}{2}\left(\frac{\partial \rho_\textrm{s}}{\partial a_\textrm{3D}^{-1}}\right)_{\mu, T, w, H }\label{rhos}.
\end{eqnarray}

The exact relations (\ref{n}-\ref{rhos}) bring new physical
insight into the correlations between the contact and other physical quantities, including the magnetization and superfluid density that characterizes magnetic and transport properties, respectively, which have not been explored in the literature. Among these exact relations,  equation (\ref{rhos}) is of particular interest. It directly correlates the contact with $\rho_\textrm{s}$ characterizing superfluid phase transitions.  Despite 
$c$ being
finite in both the normal and superfluid phases, there is a difference. Equation (\ref{rhos}) shows that in the normal phase, $\left({\partial c}/{\partial w^2}\right)_{\mu, T, H, a_\textrm{3D}^{-1}}=0$ and the contact remains unchanged after a slow rotation is turned on, since $\rho_\textrm{s}\equiv0$.  In the superfluid phase, $\rho_\textrm{s}$ in general depends on the scattering length, and therefore $\left({\partial c}/{\partial w^2}\right)_{\mu, T, H, a_\textrm{3D}^{-1}}$ is finite. In particular, in a stationary system with $w=0$,  $\rho_\textrm{s}$ follows the standard scaling law near the  transition point, $\rho_\textrm{s}=A(\delta-\delta_\textrm{c}(a^{-1}_{3D}))^{\zeta}$, where the tuning parameter $\delta$ can be $T$ and $\mu$ (or $H$) for classical and quantum phase transitions, respectively. Here, $A$ is independent of $\delta$ and $\zeta$ is the corresponding critical exponent. One then obtains the scaling law for $\left({\partial c}/{\partial w^2}\right)_{\mu, T, H, a_\textrm{3D}^{-1}}$ near the transition point, 
\begin{equation}
\left(\frac{\partial c}{\partial w^2}\right)_{\mu, T, H, a_\textrm{3D}^{-1}}\Big|_{w=0}=\frac{A\zeta}{2}\frac{\partial{\delta_\textrm{c}}}{\partial a_\textrm{3D}^{-1}}(\delta-\delta_\textrm{c})^{\zeta-1}\label{sfc}
\end{equation}
Equation (\ref{sfc}) shows that the exponent of $\left({\partial c}/{\partial w^2}\right)_{\mu, T, H, a_\textrm{3D}^{-1}}$ is entirely determined by the universality class of the superfluid phase transition. The above properties of the contact can be easily tested in experiments on trapped atoms, where the superfluid and normal phases are distributed in different regions of
the trap. With the high resolution images available in current experiments, the local contact density $c$ can be extracted as a function of $\mu$ using $c=({\partial P}/{\partial a_\textrm{3D}^{-1}} )_{\mu, T, H, w}$. One then could examine the distinct responses of $c$ to rotation, $\left({\partial c}/{\partial w^2}\right)_{\mu, T, H, a_\textrm{3D}^{-1}}$, in both the superfluid and normal phases.

Equations (\ref{n}-\ref{m}) can also be experimentally tested. Near the phase transition point, the scaling law in a system with $w=0$  
for a quantity $O$ takes the form $O=O_\textrm{r}+B_O(\delta-\delta_\textrm{c}(a_\textrm{3D}^{-1}))^{\eta_O}$, where $O=n$, $M$ or $s$, $O_\textrm{r}$ is the regular part of $O$, and $\eta_O$ is the corresponding critical exponent. One then obtains
\begin{eqnarray}
\left(\frac{\partial c}{\partial \mu}\right)_{T, H, w=0, a_\textrm{3D}^{-1}}=\frac{\partial n_\textrm{r}}{\partial a_\textrm{3D}^{-1}}-B_n\eta_n\frac{\partial{\delta_\textrm{c}}}{\partial a_\textrm{3D}^{-1}}(\delta-\delta_\textrm{c})^{\eta_n-1}\label{nc},\\
\left(\frac{\partial c}{\partial T}\right)_{\mu,  H, w=0, a_\textrm{3D}^{-1}}=\frac{\partial s_\textrm{r}}{\partial a_\textrm{3D}^{-1}}-B_s\eta_s\frac{\partial{\delta_\textrm{c}}}{\partial a_\textrm{3D}^{-1}}(\delta-\delta_\textrm{c})^{\eta_M-1}\label{mc},\\
\left(\frac{\partial c}{\partial H}\right)_{\mu, T, w=0, a_\textrm{3D}^{-1}}=\frac{\partial M_\textrm{r}}{\partial a_\textrm{3D}^{-1}}-B_M\eta_M\frac{\partial{\delta_\textrm{c}}}{\partial a_\textrm{3D}^{-1}}(\delta-\delta_\textrm{c})^{\eta_s-1}\label{sc},
\end{eqnarray}
where the subscripts on the derivatives of $O_\textrm{r}$ have been suppressed. The above differential forms also indicate scaling laws for $c$. For instance, if one chooses $\delta=\mu$, then from equation (\ref{nc}) one obtains $c=c_\textrm{r}-B_n\frac{\partial{\mu_\textrm{c}}}{\partial a_\textrm{3D}^{-1}}(\mu-\mu_\textrm{c})^{\eta_n}$, where $c_\textrm{r}=\int_{-\infty}^{\mu} d\mu'({\partial n_\textrm{r}}/{\partial a_\textrm{3D}^{-1}})$.  Note that the dependences of $c$ and $n$ on $\mu$ have the same exponent, so that one concludes that
\begin{equation}
\frac{c-c_\textrm{r}}{n-n_\textrm{r}}=-\frac{\partial{\mu_\textrm{c}}}{\partial a_\textrm{3D}^{-1}}\label{cp}.
\end{equation}
In particular, if $c_\textrm{r}=n_\textrm{r}=0$, one sees that the contact per particle in the critical region becomes a constant that is entirely determined by $-{\partial{\mu_\textrm{c}}}/{\partial a_\textrm{3D}^{-1}}$. As the non-uniform distribution of trapped atoms allows experimentalists to trace the dependence of the contact on the chemical potential \cite{Ho},  equations (\ref{nc}, \ref{cp}) can be directly tested in experiments.

\begin{figure*}[tbp]
\includegraphics[width=1.0\linewidth]{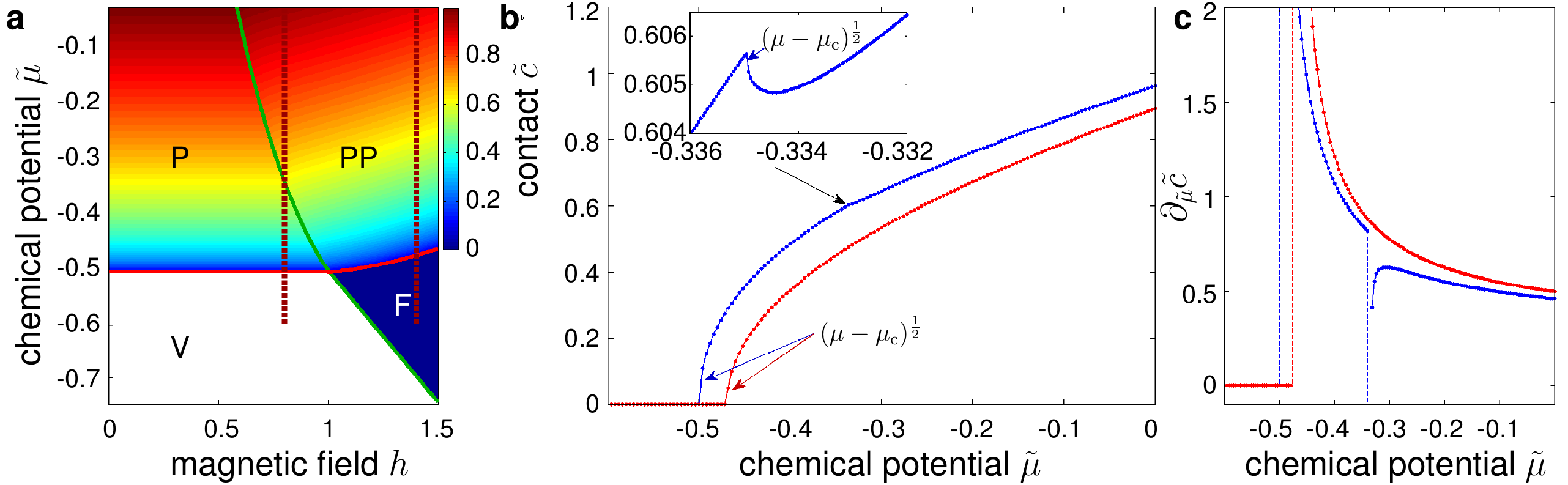}
\caption{contact of one-dimensional two-component  fermions with  zero range interaction at zero temperature. ({\bf a}) A contour plot of contact as a function of the dimensionless chemical potential  $\tilde{\mu}\equiv \mu/\epsilon_\textrm{b}$ and magnetic field $h\equiv H/\epsilon_\textrm{b}$, where $\epsilon_\textrm{b}=\hbar^2/(m a_\textrm{1D}^2)$ is the binding energy determined by the one-dimensional scattering length $a_\textrm{1D}$.  The notations V, P, F, PP stand for vacuum, fully-paired, fully-polarised and partially-polarised phase, respectively. Red and green curves represent the phase boundaries obtained from thermodynamic Bethe ansatz equations.  Vertical dashed
lines correspond to two cuts at fixed $h = 0.8$ and $h = 1.4$. ({\bf b}) contact is continuous across the quantum critical points. 
 For the transition { V-P} and { F-PP}, the dimensionless contact $\tilde{c}\equiv  {c}/{\epsilon_\textrm{b}^2}$ continuously increases from zero as $\tilde{c}\sim (\tilde{\mu}-\tilde{\mu}_\textrm{c} )^{\frac{1}{2}}$. For the transition  { P-PP},  contact is finite on both sides of the transition point and a kink exists as $\tilde{c}-\tilde{c}_\textrm{r} \sim (\tilde{\mu}-\tilde{\mu}_\textrm{c} )^{\frac{1}{2}}$,  indicating the discontinuity of the derivative of contact.  ({\bf c})  The derivative of contact with respect to $\tilde{\mu}$ becomes divergent as $\tilde{c}-\tilde{c}_\textrm{r}  \sim  (\tilde{\mu}-\tilde{\mu}_\textrm{c} )^{-\frac{1}{2}} $ in this one-dimensional system at all the transitions {V-P}, { P-PP} and { F-PP}  for fixed values of ${h} =\,0.8$ (blue line) and ${h}=\,1.4$ (red line).  }
\label{fig:phase}
\end{figure*}

{\bf Contact in an exactly solvable one-dimensional Fermi gas} 

Whereas the above discussion applies to all ultracold atomic systems, it is particularly interesting  to use an exactly solvable model to demonstrate some of the critical behaviour of contact. Here, we consider equations (\ref{nc}, \ref{cp}), since they can be implemented in experiments easily without a rotation. In one dimension, equation (\ref{3D}) becomes
\begin{equation}
dP=nd\mu+s dT+MdH-\frac{\rho_\textrm{s}}{2} dw^2-\it{c}\,da_\textrm{1D}, \label{1D}
\end{equation}
where $a_\textrm{1D}$ is the one-dimensional scattering length and $c$ differs from the ordinary definition by a trivial pre-factor $\hbar^2/(2m)$. Each of the equations (\ref{3D}-\ref{cp}) has a direct analogue in one dimension that is obtained by the simple replacement $a_\textrm{3D}^{-1}\rightarrow -a_\textrm{1D}$. We study a one-dimensional 
Fermi gas with $\delta$-function interactions, described by the Yang-Gaudin  Hamiltonian \cite{Yang,Gaudin,GBL}
\begin{eqnarray}
\mathcal{H} =-\frac{\hbar ^{2}}{2m}\sum_{i=1}^{N}\frac{\partial ^{2}}{
\partial x_{i}^{2}}+g_\textrm{1D} \sum_{i=1}^{N_{\uparrow}}\sum_{j=1}^{N_{\downarrow }}\delta \left( x_{i}-x_{j}\right)+ E_z,\label{FM}
\label{Hamiltonian}
\end{eqnarray}
where $E_z= -\frac{1}{2}HM=-\frac{1}{2}H( n_{\uparrow }-n_{\downarrow })$ is the Zeeman energy induced by a magnetic field $H$,  $g_\textrm{1D} =-\frac{2\hbar^2 }{ma_\textrm{1D}}$ characterizes the interaction strength determined by the effective  one-dimensional scattering length $a_\textrm{1D}=-a_{\perp}^2/a+A a_{\perp} $\cite{Olshanii1998}, $a_{\perp}$ is the transverse oscillation length, and $A\approx 1.0326$.  We introduce the polarization $P=(n_\ua-n_\da)/n$, and define a dimensionless interaction parameter $\gamma =mg_\textrm{1D} /(n\hbar ^{2})=-2(na_\textrm{1D})^{-1}$ for our analysis, choosing natural units $2m=\hbar =k_{B}=1$.


The model described by equation (\ref{FM}) has been  solved using the Bethe ansatz \cite{Yang,Gaudin} and has had a tremendous impact in statistical mechanics. The experimental developments in studying one-dimensional fermions \cite{Moritz:2005,Liao:2010,Zurn:2012,Wenz:2013,Pagano} have inspired significant interest in relating theoretical results to experimental observables \cite{Cazalilla,GBL}.  
It was found  \cite{Yang-Yang,Takahashi:1971,Takahashi} that, although the thermodynamic Bethe Ansatz (TBA) equations involve nontrivial collective behaviour of the particles, i.e., the motion of one particle depends on all others, the total effect of the complex behaviours of all the individual particles leads to qualitatively new forms of simplicity in many-body phenomena \cite{Zhao,Guan-Ho}.

\begin{figure}[htp]
\includegraphics[width=1\linewidth]{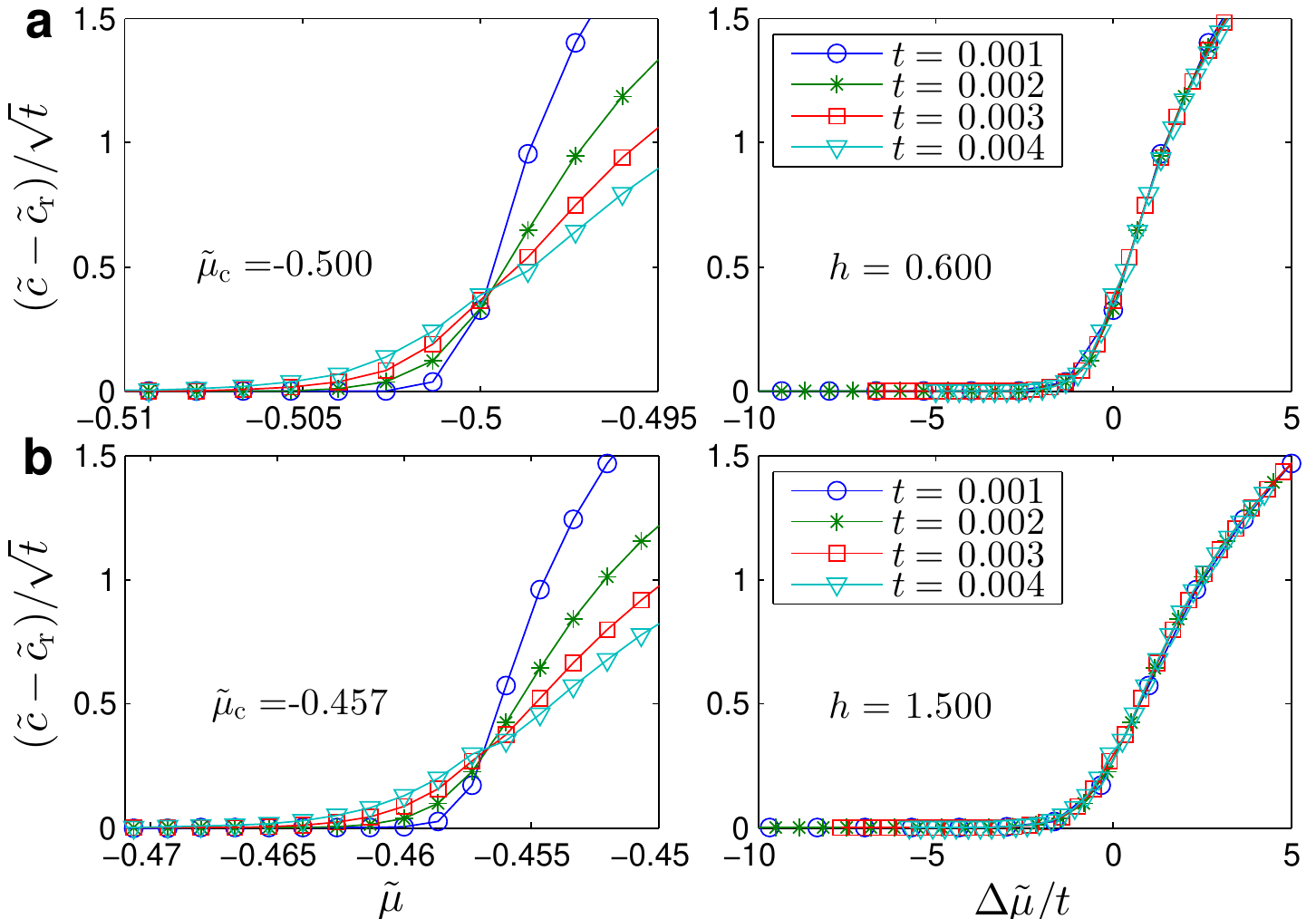}
\caption{ Scaling laws of contact determined by the universality class of the phase transition in the quantum critical region. The left panels show the temperature-scaled contact $\tilde{c}/\sqrt{t}$, where $t=T/\epsilon_\textrm{b}$ is the dimensionless temperature, as a function of the chemical potential near the phase boundaries  {V-P} ({\bf a})  and { F-PP} ({\bf b}), respectively. Curves at different temperatures intersect at the quantum critical point as predicted by  Eq. (\ref{C-form}). The right panels show that the rescaled  contact vs temperature-scaled chemical potential $\Delta \tilde{\mu}/t=\left(\tilde{\mu} -\tilde{\mu}_\textrm{c} \right)/t$ at different temperatures collapse into a single line, characteristic of critical behaviour  in the quantum critical region. These data collapses confirm the critical  dynamic exponent $z=2$ and correlation length exponent $\nu=1/2$ in terms of the universal scaling in Eq. (\ref{C-form}).}
\label{fig:contact}
\end{figure}

\begin{figure}[htbp]
\includegraphics[width=1\linewidth]{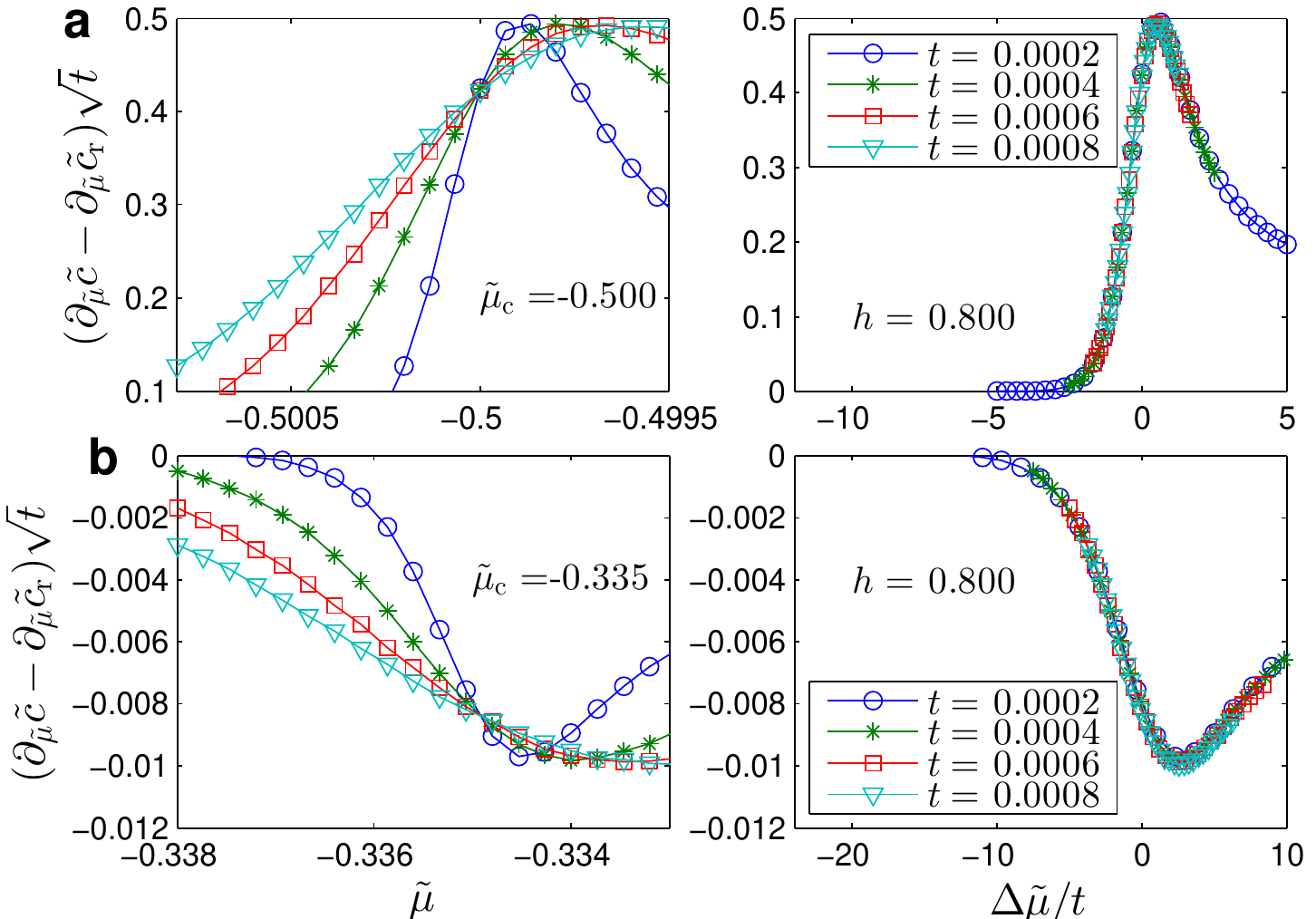}
\caption{ Scaling laws of the derivative of contact with respect to the chemical potential in the quantum critical region. The left panels show  that different curves of  $\partial\tilde{c}/\partial \tilde{\mu}$, where $\tilde{\mu}=\mu/\epsilon_\textrm{b}$ is the dimensionless chemical potential, near the transition { V-P} ({\bf a})  and { P-PP} ({\bf b}),  intersect at the quantum critical point. The right panels show the collapse of $(\partial_{\tilde{\mu}}\tilde{c})\sqrt{t}$ vs the temperature-scaled chemical potential $\Delta \tilde{\mu}/ t=\left(\tilde{\mu} -\tilde{\mu}_\textrm{c} \right) /t$  into a single curve.  Results in both panels confirm the scaling laws predicted by  Eq. (\ref{PC-form}). From the temperature-scaled contact at different temperatures, one reads off the  the critical  dynamic exponent $z=2$ and correlation length exponent $\nu=1/2$. }
\label{fig:derivative}
\end{figure}

Contact of the ground state, in the extremely polarized limit with a single spin-down atom, has been studied in \cite{Polaron,Zwerger2}. However, reaching the goal of finding critical behaviors of contact requires a theoretical framework, beyond mean-field theory, capable of analytically deriving the thermodynamic properties of such gases at finite temperatures (Supplementary Note ). 
This has been a fundamental challenge in theoretical physics, due to the strong interaction between the atoms. Here we compute the contact by numerically solving the TBA equations and obtaining analytic expressions in  the  physical regime $T\ll \epsilon_\textrm{b},H$ and $\gamma \gg 1$, where $\epsilon_\textrm{b}=2a_\textrm{1D}^{-2}$ is the binding energy of the pairs, and explore its behavior near the phase transition. 
Even though there is no finite-temperature phase transition in one dimension, there does exist a universal finite-temperature crossover which remarkably separates the low-energy critical Tomonaga-Luttinger liquid (TLL) with relativistic dispersion from the collective matter of Free Fermi criticality with non-relativistic dispersion. Moreover, quantum phase transitions between two of the following phases in this model, the vacuum phase
(V), the fully-paired phase (P), the fully-polarised phase (F) and  the partially-polarised phase (PP)
\cite{Zhao,Guan-Ho}, 
provide a precise description of the critical behaviors exhibited by contact  in  many-body systems.

The phase diagram Fig.~\ref{fig:phase} shows numerical results for the dimensionless contact density $\tilde{c}\equiv  {c}/{\epsilon_\textrm{b}^2}$ at  zero temperature as a function of the dimensionless chemical potential $\tilde{\mu}\equiv \mu/\epsilon_\textrm{b}$ and magnetic field $h\equiv H/\epsilon_\textrm{b}$, where $c$  is obtained from the TBA equations (Supplementary Note ) 
through $c=-\left(\partial P/\partial a_\textrm{1D}\right)_{\mu,H,T}$,  and $w$ has been set to be zero. Here we have chosen $\epsilon_\textrm{b}$ as the energy scale. Alternatively, one may choose the Fermi  energy $E_\textrm{F} $, which will not change the later discussion and results. Since we have chosen natural units by setting $\hbar$ and $2m$ to be 1, $c$ has the same dimension as $\epsilon_\textrm{b}^2$ so that $\tilde{c}$ as defined is dimensionless. Across the transition from V to P, the regular part $\tilde{c}_\textrm{r} \equiv 0$, since $\tilde{c}\equiv 0$ in V, and past the critical point $\tilde{c}$ continuously increases from zero as $\sim (\tilde{\mu}-\tilde{\mu}_\textrm{c} )^{1/2}$. Correspondingly, in P,  ${\partial \tilde{c}}/{\partial \tilde{\mu}} $ diverges as $(\tilde{\mu}-\tilde{\mu}_\textrm{c} )^{-1/2}$ at this transition point. The aforementioned scaling laws for $\tilde{c}$ and ${\partial \tilde{c}}/{\partial \tilde{\mu}}$ are derived directly from the zero temperature scaling law for density near this critical point, $n\sim (\mu-\mu_\textrm{c}(a_\textrm{1D}))^{1/2}$. By taking the derivative of $n$ with respect to $a_\textrm{1D}$, one sees that the critical exponents for $\tilde{c}$ and  ${\partial \tilde{c}}/{\partial \tilde{\mu}} $ are indeed $1/2$ and $-1/2$, respectively. Near the other transition point from $P$ to $PP$, $c$ also changes continuously with a kink $\tilde{c}-\tilde{c}_\textrm{r} \sim (\tilde{\mu}-\tilde{\mu}_\textrm{c} )^{1/2}$ and ${\partial \tilde{c}}/{\partial \mu} $ has the same $(\tilde{\mu}-\tilde{\mu}_\textrm{c} )^{-1/2}$ divergence.

At finite temperatures, ${\partial \tilde{c}}/{\partial \tilde{\mu}} $ no longer diverges (Supplementary Fig.1). Nevertheless, critical phenomena exist for both $\tilde{c}$ and  ${\partial \tilde{c}}/{\partial \tilde{\mu}} $ in a region expanded to finite temperatures, as is typical for quantum criticality. We work out the analytic expressions for $\tilde{c}$  and derive the scaling form for $\tilde{c}$ and its derivatives in the quantum critical region. For the physical regime $T\ll \epsilon_\textrm{b},H$ and $\epsilon_\textrm{b} \gg E_\textrm{F} $, $\tilde{c}$ is given explicitly by
\begin{equation}
\tilde{c}\approx 2\tilde{n}^\textrm{b}+\left[\left( \tilde{n}^\textrm{b} +2\tilde{n}^\textrm{u}\right) \tilde{p}^\textrm{b} +4\tilde{n}^\textrm{b}\tilde{p}^\textrm{u} \right], \label{contact}
\end{equation}
where
\begin{eqnarray}
\tilde{p}^\textrm{b} &=&-\frac{t^{\frac{3}{2}}}{2\sqrt{\pi}}f_{\frac{3}{2}}^\textrm{b} \left(1+\tilde{p}^\textrm{b} /8+2\tilde{p}^\textrm{u} \right),\nonumber\\
\tilde{p}^\textrm{u} &=&-\frac{t^{\frac{3}{2}}}{2\sqrt{2\pi}}f_{\frac{3}{2}}^\textrm{u} \left(1+2\tilde{p}^\textrm{b} \right).\nonumber
\end{eqnarray}
Here, we use the notation $f_{\ell}^\textrm{b,u} :\equiv Li_{\ell}(-e^{\tilde{A}_\textrm{b,u} /t})$, $\tilde{A}_\textrm{u} =\tilde{\mu}_\textrm{u}-2\tilde{p}^\textrm{b} -\frac{t^{\frac{5}{2}}}{2\sqrt{\pi}c^{3}}f_{\frac{5}{2}}^\textrm{b} $,  
$ \tilde{A}_\textrm{b}=2\tilde{\mu}_\textrm{b}-\tilde{p}^\textrm{b} -4\tilde{p}^\textrm{u} -\frac{t^{\frac{5}{2}}}{\sqrt{\pi}}\left( \frac{1}{16}f_{\frac{5}{2}}^\textrm{b} + \sqrt{2} f_{\frac{5}{2}}^\textrm{u} \right)$,  and $Li_{\ell}(x)=\sum_{k=1}^\infty x^k/k^{\ell}$ is the polylog function. We have  defined $t\equiv T/\epsilon_\textrm{b}$, $\tilde{n}\equiv n|a_\textrm{1D}|/2$, and $\tilde{p}^\textrm{u,b} \equiv P^\textrm{u,b}|a_\textrm{1D}|/|2\epsilon_\textrm{b}|$, where the labels $\textrm{b}$ and $\textrm{u}$ indicate if a quantity describes a property of bound states or unpaired particles.  The result (\ref{contact}) is valid for both the TLL phase and the critical region (Supplementary Fig.2).  Physically, $\tilde{p}^\textrm{u,b}$ and $\tilde{\mu}_\textrm{u,b}$ represent the  dimensionless pressure and chemical potential of unpaired fermions and bound pairs, respectively. Due to the residual interaction between them, $\tilde{p}^\textrm{b} $ and $\tilde{p}^\textrm{u} $ are correlated through the above coupled equations. 

It is interesting to note  that, apart from a small correction $O(a_\textrm{1D}^3)$, the terms within the square brackets of (\ref{contact})  give  the pressure of  the interacting system after subtracting that of an ideal gas consisting of single fermionic atoms with mass $m$ and composite atoms with mass $2m$,  namely   
\begin{equation}
\tilde{c}\approx 2 \tilde{n}^\textrm{b}- \left( \tilde{p} -\tilde{p}_0 \right), \label{contact-interaction}
\end{equation}
where up to the order of  $O(a_\textrm{1D}^3)$, the pressure is $p\approx \left( \tilde{n}^\textrm{b} _0+2\tilde{n}^\textrm{u}_0\right) \tilde{p}_0^\textrm{b} +4\tilde{n}^\textrm{b}_0\tilde{p}^\textrm{u} _0$. In these equations, $\tilde{n}_0^\textrm{b,u} =\partial \tilde{p}_0^\textrm{b,u} /\partial \mu_\textrm{b,u} $ and  $\tilde{p}_0=\tilde{p}_0^\textrm{b} +\tilde{p}_0^\textrm{u}$, with $\tilde{p}_0^\textrm{b} =-\frac{t^{3/2}}{2\sqrt{\pi }} Li_{3/2}(-e^{2\tilde{\mu}_\textrm{b}/t})$ and $\tilde{p}_0^\textrm{u} =-\frac{t^{3/2}}{2\sqrt{2\pi }} Li_{3/2}(-e^{\tilde{\mu}_\textrm{u}/t})$ (Supplementary Note ). Physically, $\tilde{p}^\textrm{u,b}_0$ represents
the pressures of free unpaired fermions or
bound pairs, as one sees clearly that their expressions are identical to those for non-interacting particles.  The term in the parentheses of equation (\ref{contact-interaction}) reveals an important  characteristic of contact in the strongly interacting region: it accounts for the interaction between bound pairs, and that between pairs and unpaired fermions,  in addition to the contribution from each pair itself. The high-order corrections to contact  from multi-body interaction  effects, i.e., scattering involving three pairs, are relatively small in the strong-coupling regime. In this regard, the two-body interaction, including both pair-pair and pair-unparied fermions scattering, are important for determining the critical behaviors of contact in a strongly interacting Fermi gas. On the other hand, in order to capture proper thermal and quantum fluctuations in the quantum critical region, the universal scaling behaviour of the contact requests such marginal contributions from those higher order corrections(Supplementary Note ).

\begin{figure}[htbp]
\includegraphics[width=1.0\linewidth]{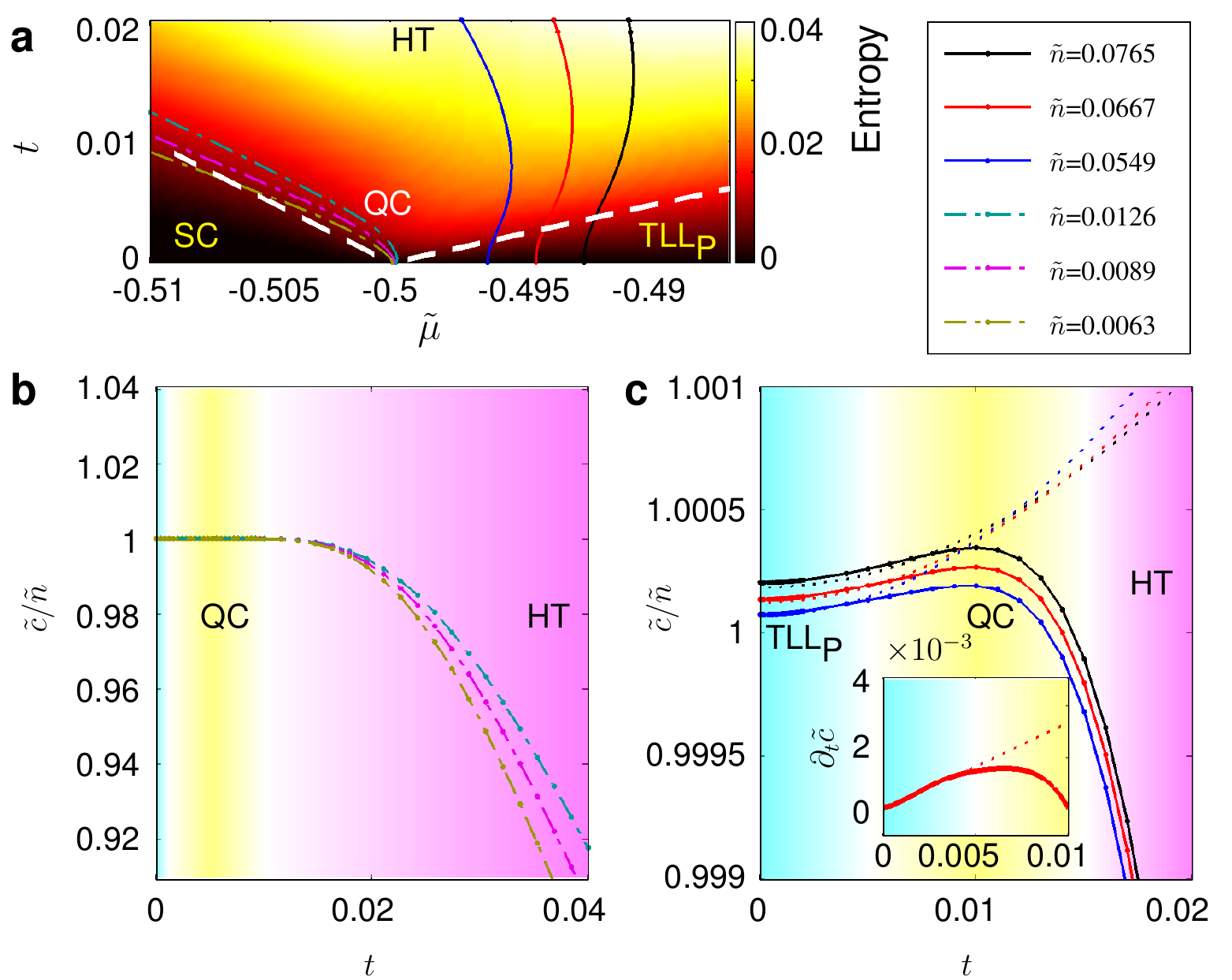}
\caption{ contact per particle at finite temperatures. ({\bf a}): Density plot of the entropy obtained from the numerical solution of thermodynamic Bethe ansatz equations for highlighting different regions  for the phase transition V-P on the $\tilde{\mu}-{t}$ plane. We denote by SC the semiclassical  region with very low density.  { QC} stands for the critical regime with non-relativistic dispersion, ${\rm TLL_p}$ is the Tomonaga-Luttinger liquid of pairs with linear relativistic dispersion, and $HT$ stands for high temperature region where universal behaviour of thermodynamics  are absent. Dashed white lines represent the crossover temperatures from QC to SC and TLL regions. ({\bf b}) contact per particle $\tilde{c}/\tilde{n}$ vs the temperature at fixed values of low densities, where $\tilde{n}=na_\textrm{1D}/2$ is the dimensionless density.  The flatness of $\tilde{c}/\tilde{n}\approx 1$  confirms the constant contact per particle as shown in equation (\ref{C-form}) in the {QC} region. ({\bf c}) $\tilde{c}/\tilde{n}$  at high densities. The solid lines show  the numerical result derived from  thermodynamic Bethe ansatz equations. The deviations of the Tomonaga-Luttinger liquid result (dotted lines)  from  thermodynamic Bethe ansatz results indicate  the breakdown of the ${\rm TLL_p}$ phase at the crossover temperature $T^*$ from ${\rm TLL_p}$ to {QC}. $T^*$ here is consistent  with the result obtained from the deviation of entropy from the linear temperature dependence of TLL. The inset shows $\partial_{t} \tilde{c}$ vs temperature, in which the deviation is more visible. A maximum of $\tilde{c}/\tilde{n}$ demonstrates the enhancement of contact when quantum effects become important in the quantum degenerate region where $t\sim \tilde{\mu}-\tilde{\mu}_\textrm{c} $. }
\label{fig:c-n}
\end{figure}

Using equation (\ref{contact}), we  find  the universal scaling form of $c$ in the quantum critical region, 
\begin{equation}
{c }={c}_\textrm{r}+\lambda_\textrm{G}T^{\frac{d}{z}+1-\frac{1}{\nu z}}\mathcal{F}(\frac{\mu-\mu_\textrm{c}(a_\textrm{1D})}{T^{\frac{1}{\nu z}}}), \label{C-form}
\end{equation}
where ${c}_\textrm{r}$ is a temperature-independent regular part,  the constant  $\lambda_\textrm{G}$  depends on $\mu_\textrm{c}$, and $\mathcal{F}(x)$ is a dimensionless scaling function which can be determined by the TBA equations (Supplementary Note ). 
The dynamic and correlation length exponents have been found to be $z=2$ and $\nu=1/2$, see the data collapse after use of scaling law (\ref{C-form}) in  Fig.~\ref{fig:contact}. From (\ref{C-form}) we obtain
\begin{equation}
\left(\frac{\partial c}{\partial \mu}\right)_{T, H, a_\textrm{1D}}=\frac{\partial c_\textrm{r}}{\partial \mu}+\lambda_\textrm{G}T^{\frac{d}{z}+1-\frac{2}{\nu z}}\mathcal{F}'(\frac{\mu-\mu_\textrm{c}(a_\textrm{1D})}{T^{\frac{1}{\nu z}}}). \label{PC-form}
\end{equation}
Fig.~\ref{fig:derivative}  shows  the scaling behaviour of this derivative of contact. Similar results can be obtained if one chooses $H$ as the tuning parameter (Supplementary Fig.3 and Supplementary Fig.4). Comparing equation (\ref{C-form}) with the standard scaling form of the density in the quantum critical region, $n=n_\textrm{r}+\tilde{B}_nT^{\frac{d}{z}+1-\frac{1}{\nu z}}\mathcal{F}\left(\frac{\mu-\mu_{c}(a_\textrm{1D})}{T^{\frac{1}{\nu z}}}\right)$, we find that $\lambda_\textrm{G}=\tilde{B}_n\partial \mu_\textrm{c}/\partial a_\textrm{1D}$ and
\begin{equation}
\frac{c-c_\textrm{r}}{n-n_\textrm{r}}=\frac{\partial \mu_\textrm{c}}{\partial a_\textrm{1D}}\label{cp1D},
\end{equation}
in analogy to equation (\ref{cp}) in three dimensions.   For the phase transition V-P, $\mu_\textrm{c}=-1/a_\textrm{1D}^2$ and ${\partial \mu_\textrm{c}}/{\partial a_\textrm{1D}}=2/a_\textrm{1D}^3$, so that equation (\ref{cp1D}) can be rewritten as $(\tilde{c}-\tilde{c}_\textrm{r} )/({\tilde{n}-\tilde{n}_\textrm{r}})=1$. For other phase transitions, such as P-PP and F-PP,  this ratio has different constant values. 
The scaling forms, equations (\ref{C-form}-\ref{PC-form}), lead to the intersection of the scaled quantities $(c-c_\textrm{r})/\sqrt{T}$ and $\sqrt{T}({\partial c}/{\partial \mu}-{\partial c_\textrm{r}}/{\partial \mu})$ for different temperatures at $\mu_\textrm{c}$ in our system. If one further plots these quantities as functions of $(\mu-\mu_\textrm{c})/T$, different curves collapse to a single one. Such intersections and data collapses are characteristic for the quantum critical behaviour  of contact. We have numerically confirmed the validity of these scaling forms for all interaction strengths. 

We now turn to the contact per particle in the quantum critical region. To highlight the quantum critical region and other ones in the $\tilde{\mu}-t$ plane, Fig.~\ref{fig:c-n}  shows a density plot of the entropy. The regions are separated by a crossover temperature $T^*$,  shown by the white dashed line in Fig.~\ref{fig:c-n}. The crossover from the quantum critical region to the TLL region, where the density is finite for $\mu>\mu_\textrm{c}$ at zero temperature, is obtained from the deviation of the entropy from the linear form of TLL \cite{GBL}.  On the other side of the transition point, the crossover temperature from the quantum critical region to the semiclassical region, where the density is exponentially small, is obtained by setting the thermal wavelength equal to the inter-particle spacing. In Fig.~\ref{fig:c-n} (a), three curves are shown for the rather small fixed values of the density $\tilde n$ listed in Fig.~\ref{fig:c-n} (b). One can see that a very large portion of the trajectory at such constant densities remains in the quantum critical region. As a result,  $\tilde{c}/\tilde{n}$ becomes 1. In Fig.~\ref{fig:c-n} (b), numerical results for the scaled contact per particle for these three densities are shown to satisfy $\tilde{c}/\tilde{n}=1+O(10^{-5})$ up to the temperature scale $t=10^{-2}$, which corresponds to a ratio of the temperature to the chemical potential ${t}/|\tilde{\mu}-\tilde{\mu}_\textrm{c} |\sim 1$. These results directly confirm equation (\ref{cp1D}). \\

At higher densities and with increasing temperature, the trajectory at constant density first enters the TLL region, quickly passes the quantum critical region, and eventually enters the high temperature region with ${t}\gg |\tilde{\mu}-\tilde{\mu}_\textrm{c} |$, where the entropy density becomes large and the universal scaling laws of contact fail, as shown in Fig.~\ref{fig:c-n} (a).  Below $T^*$, and in the TLL phase of the paired fermions, referred to as phase ${\rm TLL_p}$, the contact in the strong coupling regime is given by 
\begin{eqnarray}
\left(\frac{\tilde{c}}{\tilde{n}}\right)_\textrm{TLL}\approx 1+\frac{\pi^2\tilde{n}^3}{24}+\frac{t^2}{6\tilde{n}}.\label{c-TLL}
\end{eqnarray}
Fig.~\ref{fig:c-n} (c) shows both the numerical results of ${\tilde{c}}/{\tilde{n}}$ at large densities and the result of the TLL theory based on equation (\ref{c-TLL}). It is clear that  the growth of ${\tilde{c}}/{\tilde{n}}$ at low temperatures is described well by equation (\ref{c-TLL}). The deviation from the TLL result shows a breakdown of the TLL  at crossover temperature $T^*$.   More interestingly, one sees that before $\tilde{c}/\tilde{n}$ eventually decreases at  high temperatures, a maximum occurs around ${t}\approx 0.01$ (about $0.1-0.5T_\textrm{F}$ for strong attractive regime \cite{Liao:2010}),  which corresponds to a quantum degenerate temperature ${t}/(\tilde{\mu}-\tilde{\mu}_\textrm{c})\sim 1-2 $. Such a maximum  indicates that the contact per particle gets enlarged in the quantum degenerate region, similar to the possible enhancement of the contact near the transition temperature of three-dimensional fermions \cite{Jin3}.  \\

\noindent
{\bf Discussion}

Whereas the Tan relations have revealed how contact controls  various thermodynamic quantities, it is in general difficult to make  quantitative predictions as to how contact depends
on the many-body physics of the system.  Our results have shown that in the critical region near a phase transition point, contact and its derivatives are uniquely determined by the universality class of the phase transition. The exact thermodynamic relations shown in equations (1-5) lead to both new insights into fundamental physics and profound applications for connecting contact and macroscopic quantum phenomena.  Whereas these relations are exact for any microscopic parameters, they are particularly useful in the critical region for establishing exact relations between the universal scaling behaviors of contact and those of other thermal, magnetic and transport  quantities.  In particular,  we have  proved that contact in one dimension  not only  provides an unambiguous determination of the TLL phases and but also identifies in a novel fashion the  universality class of  quantum critical interacting many-body systems. 

Moreover, equations (1-5) can  be used to ultimately settle the aforementioned controversy over the contact of the  three-dimensional unitary Fermi gas  near the superfluid phase transition point. On the experimental side, our results suggest that high resolution in-situ images may be used to obtain precision measurements of the local pressure and contact as a function of temperature and other microscopic parameters, so that an average in the trap is not necessary. Such experiments will also be useful for exploring the size of the critical region, which is predicted to be of the order unity in the unitary limit\cite{Ed}. 
On the theoretical side, whereas a number of approaches have obtained a continuous contact across the transition point, consistent with the prediction of our exact thermodynamic relations, one needs to examine whether the results produced by a theory indeed satisfy the exact thermodynamic relations in equations (1-5).   

In this Article, we have focused on continuous phase transitions, where all physical quantities, including both superfluid density and contact, are continuous across the transition point. It is worth pointing out that a unique phase transition occurs in two-dimensional superfluids, where the superfluid density has a finite jump, and meanwhile other thermodynamic quantities remain continuous, at the Berezinskii-Kosterlitz-Thouless transition point. It would be interesting to explore whether contact could signify such a finite jump of superfluid density controlled by the deconfinement of topological excitations, i.e., vortices in two-dimensional superfluids. 

Highly controllable ultra-cold atoms are ideal platforms for exploring both universality of dilute systems governed by contact and universal critical phenomena near a phase transition point in many-body systems. In particular, current experiments with ultracold atoms are capable of measuring the critical behaviors of contact in all dimensions. We hope that our work will stimulate more studies on the intrinsic connection between these two types of fundamental phenomena on universality in physics. \\

\noindent {\bf  Method}

For the  attractive spin-1/2 Fermi gas at finite temperatures, the
thermodynamics of the homogeneous system is described by two coupled Fermi
gases of bound pairs and excess fermions in the charge sector and
ferromagnetic spin-spin interaction in the spin sector, namely the TBA equations read \cite{Takahashi}
\begin{eqnarray}
 \varepsilon^\textrm{b} &=&2(k^2-\mu_\textrm{b} )+a_2*\varepsilon^\textrm{b} _- +a_1*\varepsilon^\textrm{u} _-,\nonumber\\
 \varepsilon^\textrm{u} &=&k^2-\mu_\textrm{u} +a_1*\varepsilon^\textrm{b} _--\sum_{m=1}^\infty a_m*\varepsilon^m_-,\nonumber \\
 \varepsilon^m&=&mH+a_m*\varepsilon^\textrm{u} _-+\sum_{\ell =1}^\infty T_{m\ell }*\varepsilon^\ell _-\label{TBA}
 \end{eqnarray}
with $m=1, \ldots \infty$.  In the above equations  $*$ denotes the convolution integral, $a_m(x)=\frac{1}{2\pi}\frac{m|g_\textrm{1D} |}{(mg_\textrm{1D} /2)^2+x^2}$ and $\varepsilon_-^{\textrm{b,u},m} =-T \ln(1+e^{-\varepsilon^{\textrm{b,u},m}  /T})$.
Here $\varepsilon^{\textrm{b,u},m} $ are the dressed energies for
bound pairs,  excess single fermions and  $m$-strings of spin wave bound states,  respectively. 
These dressed energies account for excitation energies above Fermi surfaces. 
In the above equations the function $T_{m\ell}(k)$ is given by $T_{nm}(k)=A_{nm}(k)-\delta_{nm}\delta(k)$ with 
$A_{nm}=a_{|n-m|}+2a_{(|n-m|+2)}+\cdots +2a_{(n+m-2)}+a_{(n+m)}$, see \cite{Takahashi}.

The effective chemical potentials of unpaired fermions and pairs were
defined by $\mu_\textrm{u} =\mu+H/2$ and $\mu_\textrm{b} = \mu+\epsilon_\textrm{b}/2$.  
The thermal potential  per unit length $P=p^{\rm u}+p^{\rm b}$ is given in
terms of the effective pressures  $p^\textrm{u,b} =-\frac{r} {2\pi}
\int_{-\infty}^{\infty}dk\varepsilon_-^\textrm{u,b} (k)$
with  $r=1$ and $2$   for the unpaired fermions and   bound pairs.

The strategy for working out scaling form of contact near the critical points is to firstly  perform analytical  calculation of contact near different phase transition points in the physical regime $\left| \gamma \right| \gg 1$ and $t \ll 1$. 
Then we confirm the analytical result  of the  universal scaling forms by numerically solving the TBA equations of  the model for all interacting strengths.  
To this end, we first present the analytical expression of the total pressure $P=p^\textrm{b} +p^\textrm{u} $  for the regime $\left| \gamma \right| \gg 1$ and $t
\ll 1$ \cite{Guan-Ho}. 
\begin{eqnarray}
  \label{pb} p^\textrm{b}  &=& - \frac{ T^{\frac{3}{2}} }{\sqrt{2 \pi}}Li_{\frac{3}{2}}
  (- e^{A_\textrm{b}  / T}) (1 + \frac{p^\textrm{b} }{4| g_\textrm{1D}|^3} + \frac{4 p^\textrm{u} }{|g_\textrm{1D} |^3}),\label{pressure1}\\
  \label{pu} p^\textrm{u}  &=& - \frac{T^{\frac{3}{2}}}{2 \sqrt{\pi}}  Li_{\frac{3}{2}}
  (- e^{A_\textrm{u}  / T}) (1 + \frac{4 p^\textrm{b} }{|g_\textrm{1D} |^3}) \label{pressure2}
\end{eqnarray}
with the functions 
\begin{eqnarray}
  A_\textrm{b}  &=& 2 \mu + \frac{|g_\textrm{1D} |^2}{ 2} - \frac{p^\textrm{b} }{|g_\textrm{1D} |} - \frac{4 p^\textrm{u} }{|g_\textrm{1D} |}  \nonumber\\
  && -
  \frac{1}{4 \sqrt{2 \pi} |g_\textrm{1D} |^3} T^{\frac{5}{2}} Li_{\frac{5}{2}} (- e^{\frac{A_\textrm{b} }{
  T} })\nonumber \\
  &&- \frac{4}{\sqrt{\pi} |g_\textrm{1D} |^3} T^{\frac{5}{2}} Li_{\frac{5}{2}} (- e^{\frac{A_\textrm{u} }
  { T}}) \label{Ab}\\
 A_\textrm{u}  &=& \mu + \frac{H }{ 2} - \frac{2 p^\textrm{b} }{|g_\textrm{1D} |} - \frac{\sqrt{2}}{\sqrt{ \pi}
  |g_\textrm{1D} |^3} T^{\frac{5}{2}} Li_{\frac{5}{2}} (- e^{\frac{A_\textrm{b} }{T}}).   \label{Au}
\end{eqnarray}
In this model the $SU(2)$ spin degree of freedom ferromagnetically couples to the unpaired Fermi sea. 
Thus  the spin wave contributions to the function $A_\textrm{u} $   is negligible due to an exponentially small contributions at low temperatures, see \cite{Guan-Ho}.
By iteration, these  effective pressures of bound pairs and unpaired fermions   $p^\textrm{b,u}$ can be presented in  close forms.
Here a significant observation from  equations (\ref{pressure1})  and (\ref{pressure2})  is that the pressure $P$ can be written in term of a universal scaling form near the critical fields, i.e. 
\begin{eqnarray}
\tilde{P}(t,h,\tilde{\mu})=\tilde{\cal{P}}_0+t^{d/z+1}\tilde{\cal{P}}\left(\frac{\tilde{\mu}-\tilde{\mu}_\textrm{c} }{t^{1/\nu z}},\frac{h-h_c}{t^{1/\nu z}}\right),
\end{eqnarray}
where the dimensionless pressure $\tilde{P} \equiv P/|g\epsilon_\textrm{b}|$,   $\tilde{\cal{P}}_0$ is the background pressure and $\tilde{\cal{P}}$  is the dimensionless scaling function.
The dimensionless critical chemical potential $\tilde{\mu}_\textrm{c} =\mu_\textrm{c}/\epsilon_\textrm{b}$ and critical field $h_c=H_c/\epsilon_\textrm{b}$ depend on the interaction strength $g_\textrm{1D} $.  Therefore  contact would essentially  possesses universal scaling form near each critical point.  

In principle the TBA equations (\ref{TBA})  in the paper present   full thermodynamical properties of the model for  all temperature regimes and  interaction strength. 
Analytical result obtained above are useful for carrying out full thermodynamics of the model throughout all interaction regimes. 
In the present paper the numerical calculations have been performed  basing on the TBA equations  of the spin-1/2 Fermi gas with attractive interaction (\ref{TBA}).
The TBA equations (\ref{TBA})  involve infinite number of  nonlinear integral equations  accounting different lengths  of spin strings (spin wave bound states).  
This  renders one to access the thermodynamics of the model analytically and numerically. 
The key  observation is that for $n$ very large the function $a_n(x)\to 0$. 
For the string number $n$ is greater than a  critical cutoff value of the  $n_\textrm{c}$-length spin strings,  the  value  $\varepsilon^{n_\textrm{c}}$ is independent of the  interaction.
Consequently, the contributions to the $ \varepsilon^\textrm{u} $ from higher spin strings, i.e. $n>n_\textrm{c}$, can be calculated analytically. 
By iteration, one   finds  that the  value of $\varepsilon^{n}$ for $n>n_\textrm{c}$  is  the same as  the solution of the TBA equations (\ref{TBA-n}) with  $g\to \infty$, see \cite{Takahashi}
\begin{equation}
  \varepsilon^{n}(k):\equiv \varepsilon^{n,\infty}(k) = T \ln \Big[\Big( \frac{\sinh \frac{n + 1}{2 T} H}{\sinh \frac{H}{2 T}}\Big)^2 - 1\Big].   \label{eph} 
\end{equation}

In our numerical  program, we fixed  the value of $n_\textrm{c}$ until the iteration error is small enough.
In order to make a proper discretisation in the variable  space $k$, we need to find a  cutoff $k_\textrm{c}$ for  the dressed energies $\varepsilon^{n}(k)$ in spin sector.
For  $|k|\to \infty$,  we see $\varepsilon^{n}(k)\to \varepsilon^{n,\infty}$
which  is  given  in eq. (\ref{eph}), while for the charge sector $\varepsilon^{\rm u,\infty}=\varepsilon^{\rm b,\infty}=\infty$.
Therefore for  $|k|>k_\textrm{c}$, we use  this constant dressed energy $ \varepsilon^{n,\infty}$  for  numerical calculation.
There exists an  error in comparison with the real dressed energies which is not flat in this region $|k|>k_\textrm{c}$.
In our  program, we also fix  the value $k_\textrm{c}$ until the iteration error is negilible.

For an arbitrary
interaction strength, we are able to truncate infinite number of
strings TBA equations to finite number of TBA equations  in terms of the variables $\varepsilon^\textrm{b,u}  =T \ln \xi^\textrm{b,u} $ and $ \varepsilon^n=T\ln \eta_n$
\begin{eqnarray}
\ln\xi^\textrm{b} (k)&=&\frac{2(k^2-c^2/4-\mu)}{T}+a_2*\ln(1+{\xi^\textrm{b} (k)}^{-1})\nonumber\\
&& +a_1*\ln\left(1+{\xi^\textrm{u} (k)}^{-1}\right),\nonumber\\
\ln\xi^\textrm{u} (k)&=&S*\ln(1+\xi^\textrm{b} (k))-S*\ln\left(1+\eta_1(k)\right), \nonumber\\
\ln\eta_1(\lambda) &=&S*\ln(1+{\xi^\textrm{u} (\lambda)}^{-1})+S*\ln(1+\eta_2(\lambda)), \nonumber\\
\ln\eta_2(\lambda) &=&S*\ln(1+\eta_3(\lambda))+S*\ln(1+\eta_2(\lambda)), \nonumber\\
& \ldots &
\nonumber
\\
\ln\eta_{n_\textrm{c}}(\lambda) &=&2S*\ln\left(\cosh(\frac{H}{2T})\sqrt{1+\eta_{n_\textrm{c}}(\lambda) }\right. \nonumber \\
&&\left.  +\sqrt{1+\sinh^2(\frac{H}{2T})(1+\eta_{n_\textrm{c}}(\lambda))}\right)\nonumber\\
&&+S*\ln(1+\eta_{n_\textrm{c}-1}(\lambda)).\label{TBA-n}
\end{eqnarray}
Here the functions $S(x)=\frac{1}{2|c|\cosh(\frac{\pi x}{|g_\textrm{1D} |})}$.  
From the parameters $\xi^\textrm{b} (k)$ and $\xi^\textrm{u} (k)$, we can get the pressures  $p^\textrm{b,u} $. 
This new set of the TBA equations provide numerical access to the  full thermodynamics of the model, including the Tomonaga-Luttinger liquid physics, quantum criticality, thermodynamics  and zero temperature phase diagram. \\\\

{\bf Acknowledgments.} 
XWG  thanks  R. Hulet, J. H. H. Perk, D. Ridout  and H.-W. Xiong for helpful discussions.
This  work has been supported by  the National Basic Research Program of China under Grant No. 2012CB922101 and NNSFC under grant numbers 11374331 and 11304357. 
The work of XWG  has been partially supported by the Australian Research Council. XWG thanks Chinese University of Hong Kong for kind hospitality.  QZ is supported by NSFC-RGC(N-CUHK453/13) and CUHK direct grant (4053083). 

{\bf Author Contributions}
Q. Z. and X.W. G. conceived the project. Y.Y. C. and Y.Z. J. performed the numerical and analytical studies on the one-dimensional model. Q. Z and X.W. G. wrote the paper. 

{\bf  Competing financial interests}  The authors declare no competing financial interests.

{\bf  Corresponding author} Correspondence should be addressed to Qi Zhou (qizhou@phy.cuhk.edu.hk) and Xi-Wen Guan (xiwen.guan@anu.edu.au).

\onecolumngrid

\newpage

\newpage

\noindent {\bf \Large Supplementary Note 1}

\vspace{0.15in}

\noindent {\bf  \large Tan's contact}

By definition  of Tan's contact $c$ 
\begin{equation}
  c = - \frac{g^2}{2} ( \frac{\partial P}{\partial g})_{\mu, H, T}
\end{equation}
and iterating the equations (20)-(23) in the main text  we  obtain 
\begin{eqnarray}
  \widetilde{c} & = & - \frac{1}{\sqrt{\pi}}
  t^{\frac{1}{2}} f_{1 / 2}^{A_\textrm{b}} - \frac{1}{2 \pi} t (f_{1 / 2}^{A_\textrm{b}})^2 -
  \frac{1}{\sqrt{2} \pi} tf_{1 / 2}^{A_\textrm{u}} f_{1 / 2}^{A_\textrm{b}} - \frac{1}{4 \pi^{3
  / 2}} t^{\frac{3}{2}} (f_{1 / 2}^{A_\textrm{b}})^3 \nonumber\\
  &  &-\frac{5}{2 \sqrt{2} \pi^{3 / 2}}
  \left( f_{1 / 2}^{A_\textrm{b}} \right)^2 f_{1 / 2}^{A_\textrm{u}}  - \frac{1}{8 \pi^2} t^2 (f_{1 / 2}^{A_\textrm{b}})^4 - \frac{9}{4 \sqrt{2}
  \pi^2} t^2 \left( f_{1 / 2}^{A_\textrm{b}} \right)^3 f_{1 / 2}^{A_\textrm{u}} \nonumber\\
  &  & -  \frac{1}{\pi^2} t^2 \left( f_{1 / 2}^{A_\textrm{b}} \right)^2 \left( f_{1 / 2}^{A_\textrm{u}}
  \right)^2 + \frac{7}{16 \pi} t^2 f_{1 / 2}^{A_\textrm{b}} f_{3 / 2}^{A_\textrm{b}} + \frac{1}{\sqrt{2} \pi} t^2 f_{1 / 2}^{A_\textrm{u}} f_{3 / 2}^{A_\textrm{b}} \nonumber\\
  &&+
  \frac{3}{\sqrt{2} \pi} t^2 f_{1 / 2}^{A_\textrm{b}} f_{3 / 2}^{A_\textrm{u}} + O \left(
  t^{\frac{5}{2}} \right). \label{G2}
\end{eqnarray}
Here we denote the dimensionless contact $\tilde{c}=c/\epsilon_\textrm{b}^2$  and  $f_n^x= Li_n (- e^{x / T})$.
The above
equation of contact looks  very complex. 
Nevertheless, the universal scaling form of contact is hidden in such complexity of this kind. 
In the Supplementary  Figure~ \ref{Supplementary-fig:C-mu}  shows a 3D contour plot $\tilde{c}/\tilde{n}$ against dimensionless temperature $t$ and chemical potential $\tilde{\mu}$ at $h=0.8$. Near the lower critical point $\tilde{\mu}_\textrm{c}=-0.5$, the flatness of $\tilde{c}/\tilde{n}$ is the  consequence of the criticality of the model as discussed in the main paper. 
The values of $\tilde{c}/\tilde{n}$ drops  very faster for the chemical potential excesses  the upper critical point $\tilde{\mu}_\textrm{c}=-0.335$ due to the increase of the polarization. 
We will present  further discussions on the critical behaviour of contact in the following part.

The derivatives of contact  connect various thermal and magnetic  properties such as density, magnetization and entropy
\begin{eqnarray}
\frac{1}{\epsilon_\textrm{b}}  \left( \frac{\partial c}{\partial \mu} \right)_{g, H, T} &=& -\left(
\frac{\partial n}{\partial g} \right)_{\mu, H, T},\\
\frac{1}{\epsilon_\textrm{b}} \left( \frac{\partial c}{\partial H} \right)_{g, \mu, T} &=&- \left(
  \frac{\partial m}{\partial g} \right)_{\mu, H, T},\\
\frac{1}{\epsilon_\textrm{b}}   \left( \frac{\partial c}{\partial T} \right)_{g, \mu, H} &=&- \left(
  \frac{\partial s}{\partial g} \right)_{\mu, H, T}.
\end{eqnarray}
We can  analytically calculate  these derivatives, namely 
\begin{eqnarray}
   \partial_{\tilde{\mu}} \widetilde{c} & = & -
  \frac{2}{\sqrt{\pi}} t^{- \frac{1}{2}} f_{- 1 / 2}^{A_\textrm{b}} - \frac{3}{\pi}
  f_{1 / 2}^{A_\textrm{b}} f_{- 1 / 2}^{A_\textrm{b}} - \frac{2 \sqrt{2}}{\pi} f_{1 / 2}^{A_\textrm{u}}
  f_{- 1 / 2}^{A_\textrm{b}} - \frac{1}{\sqrt{2 \pi}} f_{1 / 2}^{A_\textrm{b}} f_{- 1 / 2}^{A_\textrm{u}}
  \nonumber\\
  &  & + t^{\frac{1}{2}} [- \frac{3}{\pi^{3 / 2}} (f_{1 / 2}^{A_\textrm{b}})^2 f_{- 1
  / 2}^{A_\textrm{b}} - \frac{9 \sqrt{2}}{\pi^{3 / 2}} f_{1 / 2}^{A_\textrm{b}} f_{1 / 2}^{A_\textrm{u}}
  f_{- 1 / 2}^{A_\textrm{b}} - \frac{1}{\pi^{3 / 2}} (f_{1 / 2}^{A_\textrm{u}})^2 f_{- 1 /
  2}^{A_\textrm{b}} \nonumber\\
  &  & - \frac{9}{2 \sqrt{2} \pi^{3 / 2}} (f_{1 / 2}^{A_\textrm{b}})^2 f_{- 1 /
  2}^{A_\textrm{u}}] + \frac{1}{2} t [- \frac{5}{\pi^2} (f_{1 / 2}^{A_\textrm{b}})^3 f_{- 1 /
  2}^{A_\textrm{b}} - \frac{30 \sqrt{2}}{\pi^2} (f_{1 / 2}^{A_\textrm{b}})^2 f_{1 / 2}^{A_\textrm{u}}
  f_{- 1 / 2}^{A_\textrm{b}} \nonumber\\
  &  & - \frac{27}{\pi^2} f_{1 / 2}^{A_\textrm{b}} (f_{1 / 2}^{A_\textrm{u}})^2 f_{- 1 /
  2}^{A_\textrm{b}} - \frac{9}{2 \sqrt{2} \pi^2} (f_{1 / 2}^{A_\textrm{b}})^3 f_{- 1 / 2}^{A_\textrm{u}}
  - \frac{6 \sqrt{2}}{\pi^2} (f_{1 / 2}^{A_\textrm{b}})^2 f_{- 1 / 2}^{A_\textrm{u}} \nonumber\\
  &  & - \frac{6}{\pi^2} (f_{1 / 2}^{A_\textrm{b}})^2 f_{1 / 2}^{A_\textrm{u}} f_{- 1 /
  2}^{A_\textrm{u}} + \frac{7}{4 \pi} (f_{1 / 2}^{A_\textrm{b}})^2 + \frac{5 \sqrt{2}}{\pi} f_{1
  / 2}^{A_\textrm{b}} f_{1 / 2}^{A_\textrm{u}} + \frac{2}{\pi} f_{- 1 / 2}^{A_\textrm{b}} f_{3 / 2}^{A_\textrm{b}}
  \nonumber\\
  &  & + \frac{\sqrt{2}}{\pi} f_{- 1 / 2}^{A_\textrm{u}} f_{3 / 2}^{A_\textrm{b}} + \frac{8
  \sqrt{2}}{\pi} f_{- 1 / 2}^{A_\textrm{b}} f_{3 / 2}^{A_\textrm{u}}] + O \left( t^{\frac{3}{2}}
  \right), \label{dG2u}
  \end{eqnarray}
  \begin{eqnarray}
 \partial_h \widetilde{c} & = & - \frac{1}{\sqrt{2}
  \pi} f_{1 / 2}^{A_\textrm{u}} f_{- 1 / 2}^{A_\textrm{b}} - \frac{1}{2 \sqrt{2} \pi} f_{1 /
  2}^{A_\textrm{b}} f_{- 1 / 2}^{A_\textrm{u}} + \frac{1}{\sqrt{2}} t^{\frac{1}{2}} [-
  \frac{3}{2 \pi^{3 / 2}} f_{1 / 2}^{A_\textrm{b}} f_{1 / 2}^{A_\textrm{u}} f_{- 1 / 2}^{A_\textrm{b}}
  \nonumber\\
  &  & - \frac{1}{\sqrt{2} \pi^{3 / 2}} (f_{1 / 2}^{A_\textrm{u}})^2 f_{- 1 / 2}^{A_\textrm{b}}
  - \frac{5}{4 \pi^{3 / 2}} (f_{1 / 2}^{A_\textrm{b}})^2 f_{- 1 / 2}^{A_\textrm{u}}] +
  \frac{1}{2} t [- \frac{3}{\sqrt{2} \pi^2} (f_{1 / 2}^{A_\textrm{b}})^2 f_{1 /
  2}^{A_\textrm{u}} f_{- 1 / 2}^{A_\textrm{b}} \nonumber\\
  &  & - \frac{15}{2 \pi^2} f_{1 / 2}^{A_\textrm{b}} (f_{1 / 2}^{A_\textrm{u}})^2 f_{- 1 /
  2}^{A_\textrm{b}} - \frac{9}{4 \sqrt{2} \pi^2} (f_{1 / 2}^{A_\textrm{b}})^3 f_{- 1 / 2}^{A_\textrm{u}}
  - \frac{3}{\pi^2} (f_{1 / 2}^{A_\textrm{b}})^2 f_{1 / 2}^{A_\textrm{u}} f_{- 1 / 2}^{A_\textrm{u}}
  \nonumber\\
  &  & + \frac{3}{\sqrt{2} \pi} f_{1 / 2}^{A_\textrm{b}} f_{1 / 2}^{A_\textrm{u}} +
  \frac{1}{\sqrt{2} \pi} f_{- 1 / 2}^{A_\textrm{u}} f_{3 / 2}^{A_\textrm{b}} +
  \frac{\sqrt{2}}{\pi} f_{- 1 / 2}^{A_\textrm{b}} f_{3 / 2}^{A_\textrm{u}}] + O \left(
  t^{\frac{3}{2}} \right).  \label{dG2h}
\end{eqnarray}
Again, we can work out the scaling functions of these derivatives directly from the above equations. 
In Supplementary Figure~\ref{Supplementary-fig:sup-C-h}, we plot the derivative of contact $\partial \tilde{c}/\partial \tilde{ \mu}$ against chemical potential  $\tilde{\mu}$.
It is clearly see that the derivative of contact evolve into a sharp peak at the critical point. 

\vspace{0.2in}

\noindent {\bf  \large Universal Scaling Forms}

Quantum phase transitions occur at absolute
zero temperature as the driving parameters $\mu$ and $H$  are varied across the phase boundaries. 
The phase transitions are driven by quantum fluctuations with quantum critical points governed by divergent correlation lengths. 
Near a quantum critical point, the many-body system is expected to show universal scaling behaviour in the thermodynamic quantities.
 In the critical regime, a universal and scale-invariant description of the system is expected through the power-law scaling of the thermodynamic properties 
 \cite{Fisher,sachdev}.
Quantum phase transitions are uniquely characterized by the critical exponents
depending only on the dimensionality and symmetry of the system. 
In order to work out the connection of  Tan's  contact  to the criticality of the model, we first present   the dimensionless 
functions

\begin{eqnarray}
 \tilde{A}_\textrm{u} &=& A_\textrm{u} / \epsilon_\textrm{b} = \tilde{\mu} + h / 2 +
  \frac{1}{\sqrt{\pi}} t^{\frac{3}{2}} f_{3 / 2}^{\tilde{A}_\textrm{b}},  \label{dAu}\\
  \tilde{A}_\textrm{b} &=& A_\textrm{b} / \epsilon_\textrm{b} = 2 \tilde{\mu} + 1 + \frac{1}{2
  \sqrt{\pi}} t^{\frac{3}{2}} f_{3 / 2}^{\tilde{A}_\textrm{b}} + \frac{\sqrt{2}}{\sqrt{\pi}}
  t^{\frac{3}{2}} f_{3 / 2}^{\tilde{A_\textrm{u}}}. \label{dAb} 
\end{eqnarray}
From equations(\ref{dAu}) and
(\ref{dAb}), we could expand contact   (\ref {G2}) in the critical regime, i.e. $\left| \tilde{\mu} - \tilde{\mu}_\textrm{c}
\right| \ll 1$ and $\left| \tilde{\mu} - \tilde{\mu}_\textrm{c} \right| > t$  near different quantum phase transitions.

{V-P:} From vacuum $V$  to the fully-paired phase P,  the critical point is
$\tilde{\mu}_\textrm{c} = - 1 / 2, h < 1$. Taking low temperature limit near the critical point,  we can
obtain
\begin{equation}
   \tilde{A}_\textrm{u} \approx ( \tilde{\mu} - \tilde{\mu}_\textrm{c})+(h - 1) / 2, \hspace{1em} \tilde{A}_\textrm{b}
  \approx 2 ( \tilde{\mu} - \tilde{\mu}_\textrm{c}), \label{Avp}
\end{equation}
 Substituting  Eq.(\ref{Avp}) into Eq.(\ref{G2}), we can obtain the scaling forms of contact and its derivative with respect to $\mu$.
\begin{eqnarray}
 \widetilde{c} &=& - \frac{1}{\sqrt{\pi}}
  t^{\frac{1}{2}} Li_{\frac{1}{2}} (- e^{\frac{2 ( \tilde{\mu} -
  \tilde{\mu}_\textrm{c})}{t}}),   \label{G2vp}\\
  \partial_{\tilde{\mu}} \widetilde{c} &=& -
  \frac{2}{\sqrt{\pi}} t^{- \frac{1}{2}} Li_{- \frac{1}{2}} (- e^{\frac{2 (
  \tilde{\mu} - \tilde{\mu}_\textrm{c})}{t}}).\nonumber
\end{eqnarray}
In this phase $h_\textrm{c}$ is the constant. Therefore there does not exist scaling form of the derivative respect to $H$, i.e. 
$  \partial_h \widetilde{c}\approx 0 $.

{V-F:} From the vacuum V to the fully-polarized phase F the critical point is
$\tilde{\mu}_\textrm{c} = - h / 2, h > 1$. Near the critical point, we have 
obtain
\begin{equation}
  \label{Avf} \tilde{A}_\textrm{u} \approx \tilde{\mu} - \tilde{\mu}_\textrm{c}, \hspace{1em}
  \tilde{A}_\textrm{b} \approx 2( \tilde{\mu} - \tilde{\mu}_\textrm{c})+ 1 - h
\end{equation}
By expansion of Eq.(\ref{G2}) within the critical regime,  the scaling form of contact is almost zero, i.e. $  \widetilde{c} = - \frac{1}{\sqrt{\pi}}
  t^{\frac{1}{2}} Li_{\frac{1}{2}} (- e^{\frac{1 - h}{t}}) \sim 0$. This regime does not exhibit universal scaling behaviour.

{F-PP:} From the fully-polaized phase F to the partially-polarized phase PP,  the critical point is
$\tilde{\mu}_\textrm{c} = - 1 / 2 + \frac{4}{3 \pi}  (h - 1)^{3 / 2}$ and $ h>1$.
Omitting the higher order contributions from $t$ and $\tilde{\mu} - \tilde{\mu}_\textrm{c}$ we can
obtain
\begin{equation}
  \label{Afpp} \tilde{A}_\textrm{u} \approx ( \tilde{\mu} - \tilde{\mu}_\textrm{c})+ a / 2, \hspace{1em} \tilde{A}_\textrm{b} \approx 2 (
  \tilde{\mu} - \tilde{\mu}_\textrm{c})
\end{equation}
where $a = (h - 1)  (1 + \frac{2}{3 \pi}  \sqrt{h - 1})$.  
Substituting 
Eq.(\ref{Afpp}) into Eq.(\ref{G2}),  we can get the scaling forms 
\begin{eqnarray}
 \widetilde{c} &=& - \frac{1}{\sqrt{\pi}}
  t^{\frac{1}{2}} Li_{\frac{1}{2}} (- e^{\frac{2 ( \tilde{\mu} -
  \tilde{\mu}_\textrm{c})}{t}}) (1 - \frac{1}{\pi} a^{1 / 2} + \frac{1}{\pi} a^{3 / 2}),  \label{G2fpp} \\
  \partial_{\tilde{\mu}} \widetilde{c} &=&  t^{- \frac{1}{2}} Li_{-
  \frac{1}{2}} (- e^{\frac{2 ( \tilde{\mu} - \tilde{\mu}_\textrm{c})}{t}}) (-
  \frac{2}{\sqrt{\pi}} - \frac{4}{\pi^{3 / 2}} a^{1 / 2} + \frac{2}{\pi^{5 /
  2}} a).
\end{eqnarray}

{P-PP:} Similar calculations can be carried out for the phase transitions from  the phase P into phase PP,  the critical point is
$\tilde{\mu}_\textrm{c} = - h / 2 + \frac{4}{3 \pi}  (1 - h)^{3 / 2}$ and $h<1$.
Thus near the critical pint, we have 
\begin{equation}
  \label{Appp} \tilde{A}_\textrm{u} \approx \tilde{\mu} - \tilde{\mu}_\textrm{c}, \hspace{1em}
  \tilde{A}_\textrm{b} \approx 2(\tilde{\mu} - \tilde{\mu}_\textrm{c})+ b,
\end{equation}
where $b = (1 - h)  (1 + \frac{2}{\pi}  \sqrt{1 - h})$.  Substituting 
Eq.(\ref{Appp}) into Eq.(\ref{G2}), we obtain the scaling forms 
\begin{eqnarray}
 \widetilde{c} &=& \tilde{c}_0 + t^{\frac{1}{2}}\lambda
  Li_{\frac{1}{2}} (- e^{\frac{\tilde{\mu} - \tilde{\mu}_\textrm{c}}{t}}), \label{G2ppp}\\
   \partial_{\tilde{\mu}} \widetilde{c} &=& \tilde{c}_{d0} + t^{- \frac{1}{2}}\lambda_{\mu} Li_{-
  \frac{1}{2}} (- e^{\frac{\tilde{\mu} - \tilde{\mu}_\textrm{c}}{t}}).
  \end{eqnarray}
Where the constants are given by 
\begin{eqnarray}
\tilde{c}_0 &=& \frac{2}{\pi} b^{1 / 2} -
  \frac{2}{\pi^2} b+ \frac{2}{\pi^3} b^{3 / 2},\,
  \lambda=
  \frac{\sqrt{2}}{\pi^{3 / 2}} b^{1 / 2} - \frac{5 \sqrt{2}}{\pi^{5 / 2}} b +
  \frac{9 \sqrt{2}}{\pi^{7 / 2}} b^{3 / 2} - \frac{2 \sqrt{2}}{3 \pi^{3 / 2}}
  b^{3 / 2},\nonumber\\
  {c}_{d0}&=& \frac{2}{\pi} b^{- 1 / 2}
  + \frac{6}{\pi^2} + \frac{12}{\pi^3} b^{1 / 2},\,
  \lambda_{\mu} = 
  \frac{\sqrt{2}}{\pi} b^{1 / 2} - \frac{9 \sqrt{2}}{\pi^{5 / 2}} b.\nonumber
  \end{eqnarray}
  
In general, at quantum criticality, the above results can be cast into the  universal scaling
forms
\begin{eqnarray}
 \widetilde{c} &=& \widetilde{c}_0 + \lambda
  t^{\left( d / z \right) + 1 - \left( 1 / \nu z \right)}  \mathcal{F} \left(
  \frac{\tilde{\mu} -\tilde{ \mu} _\textrm{c}}{t^{1 / \nu z}} \right), \label{sG2} \\
    \partial_{\tilde{\mu}} \widetilde{c}& =&
  \widetilde{c}_{d 0} + \lambda_\mu t^{\left( d / z
  \right) + 1 - \left( 2/ \nu z \right)}  \mathcal{G} \left( \frac{\tilde{\mu} -
 \tilde{ \mu}_\textrm{c}}{t^{1 / \nu z}} \right),\label{sderivative}
\end{eqnarray}
where the scaling functions read off 
 the critical  dynamic exponent $z = 2$, correlation exponent $\nu = 1 / 2$ for
contact and its derivatives. 
In the above equations $ \widetilde{c}_{0,d 0}$, $\lambda$  and $\lambda_{\mu}$ are constants. They are  independent of the temperature. 
$ \mathcal{F} (x),  \, \mathcal{G}(x)$ are universal dimensionless scaling functions. 
Despite  the analytic results were  derived
for the strong atrractive case,  the criticality is avaliable for all interaction strength.
This nature is numerically confirmed in the main  paper.  
%


\begin{figure}[h]
  {\resizebox{9cm}{!}{\includegraphics{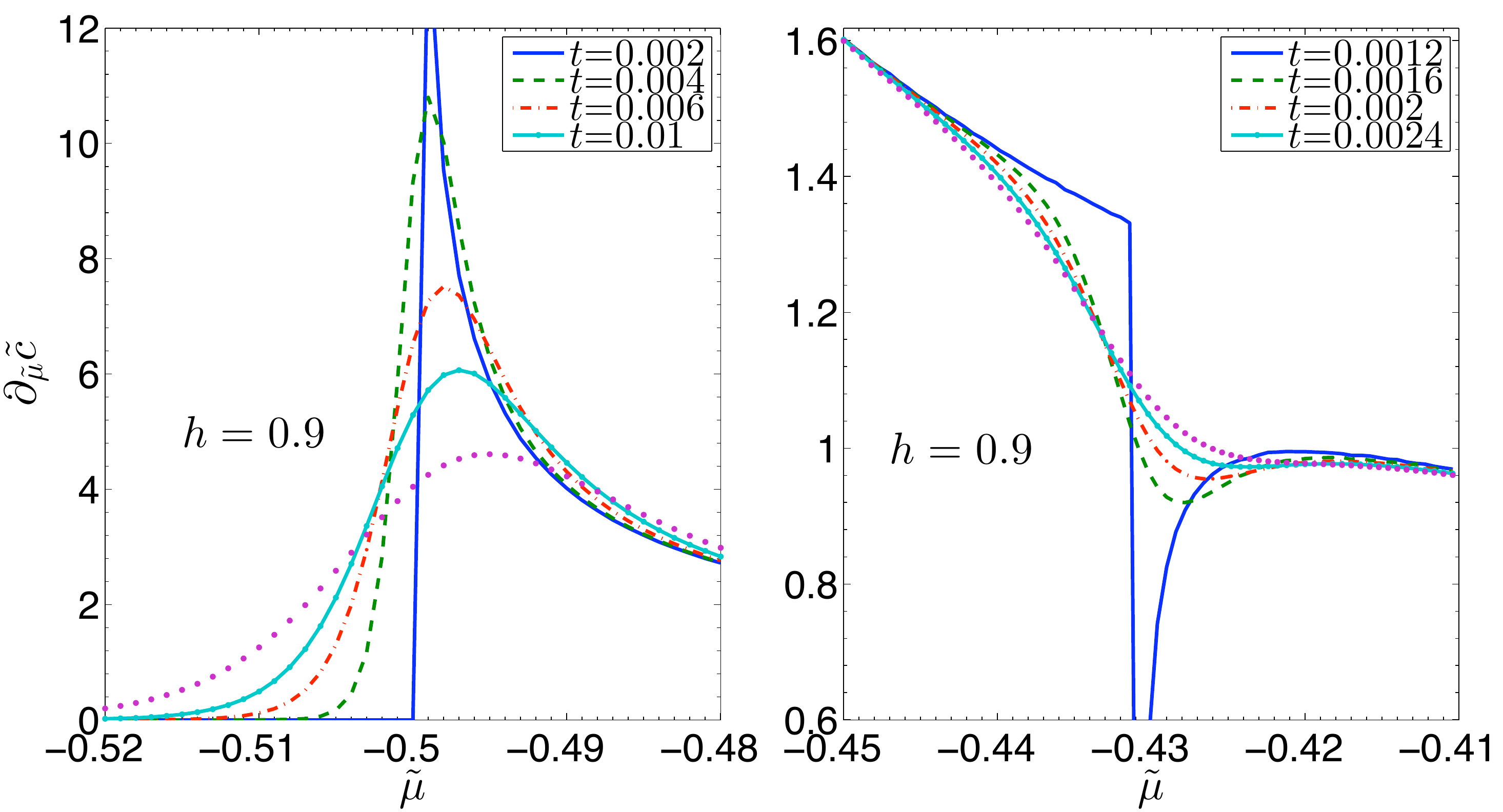}}}
  \renewcommand{\figurename}{Supplementary Figure} 
  \caption{Derivative of contact with respect to  chemical potential  vs $\tilde{\mu}$   near the phase transitions  V-P  (left panel) and P-PP (right  panel) at different temperatures.  Derivative of contact becomes divergent at $T=0$ across  these two critical points. At finite temperatures, such a divergence no longer exists.  Here the critical chemical potentials $\tilde{\mu}_\textrm{c}=-0.5$ and $\tilde{\mu}_\textrm{c}=-0.431$ for V-P and P-PP phase transitions.}
  \label{Supplementary-fig:sup-C-h}
\end{figure}

\begin{figure}[h]
  {\resizebox{9cm}{!}{\includegraphics{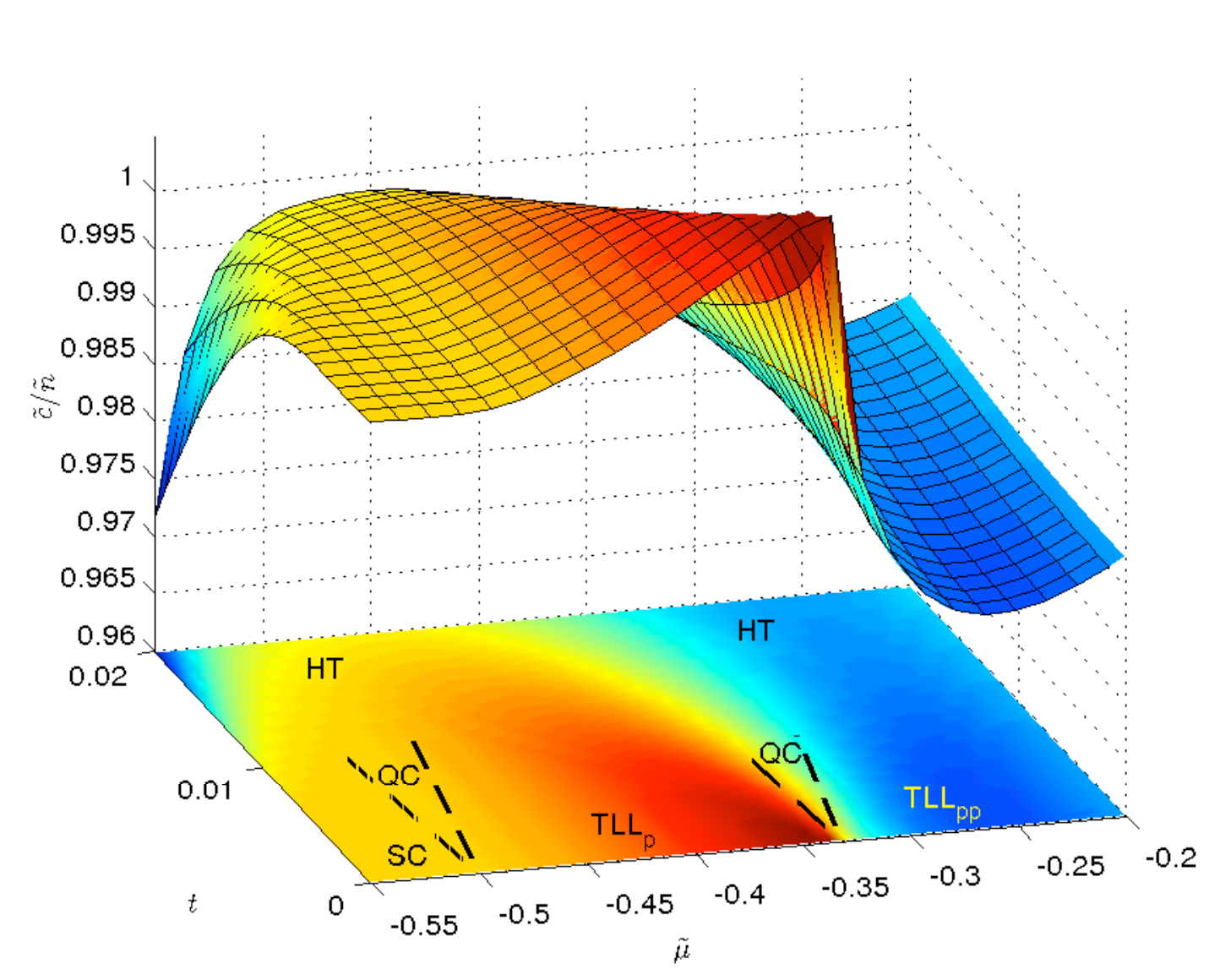}}}
    \renewcommand{\figurename}{Supplementary Figure} 
  \caption{A three-dimensional contour plot  $\tilde{c}/\tilde{n}$ against $t$ and $\tilde{\mu}$ at a fixed value of $h=0.8$. Near two critical points $\tilde{\mu}_\textrm{c1}=-0.5$ and $\tilde{\mu}_\textrm{c2}=-0.335$, different scaling behaviour are visible. See the main text about the critical behaviors of contact.    }
  \label{Supplementary-fig:C-mu}
\end{figure}

\begin{figure}[h]
  \includegraphics[scale=0.38]{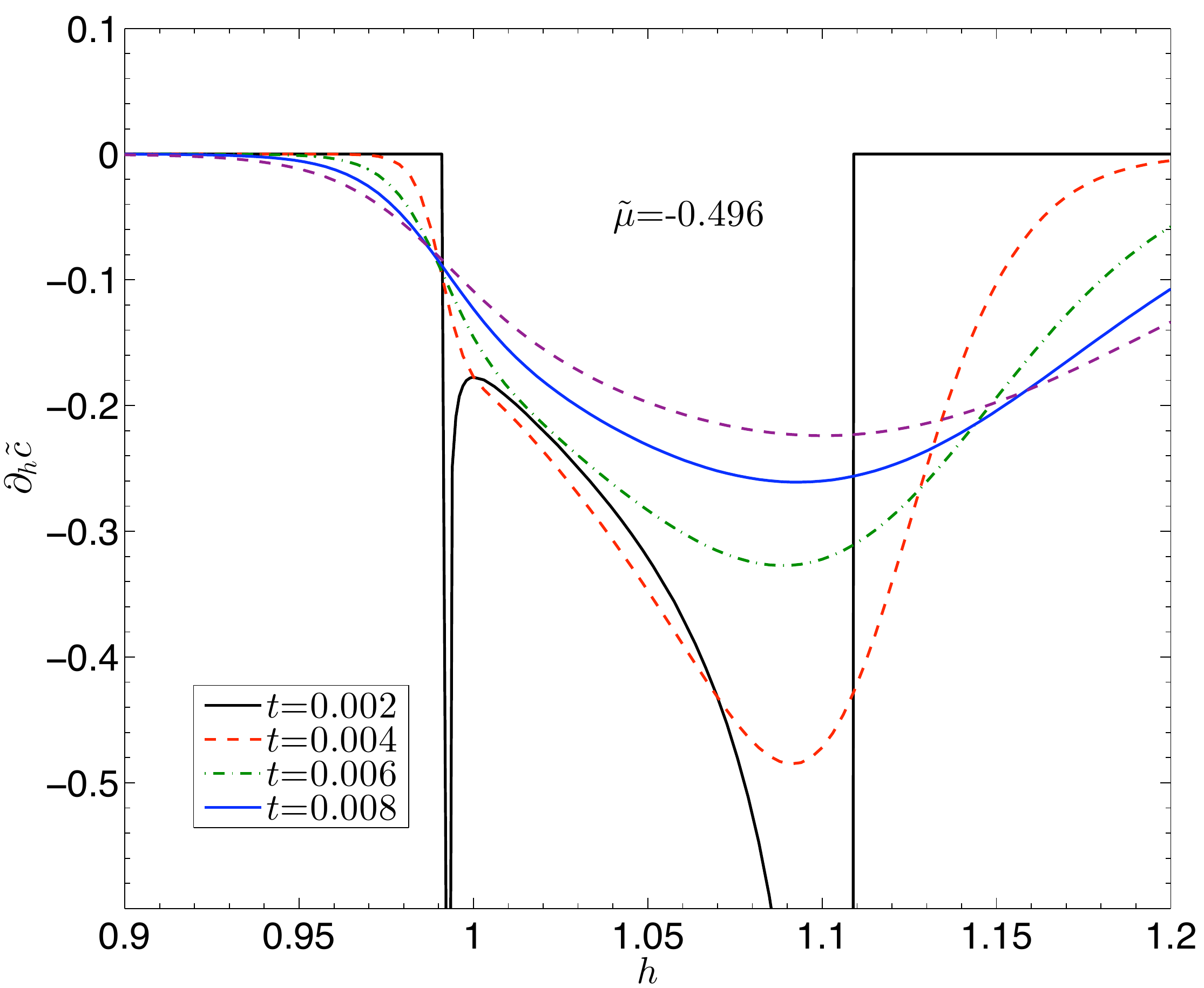} 
    \renewcommand{\figurename}{Supplementary Figure} 
      \caption{ Derivative of  contact $\partial_h\tilde{c}$ vs $h$.  Similar to $\partial_{\tilde{\mu}}{\tilde{c}}$, $\partial_{h}{\tilde{c}}$ also becomes divergent at $T=0$ across the phase transition points. For a fixed value of $\tilde{\mu}= -0.496$ the first (second)  divergent peak  presents the critical behaviour  of the gas for the phase transitions  from P to PP and  {F} to PP, respectively.}
    \label{Supplementary-fig:c-h-d}
    \end{figure}
	
\begin{figure}
	\includegraphics[scale=0.38]{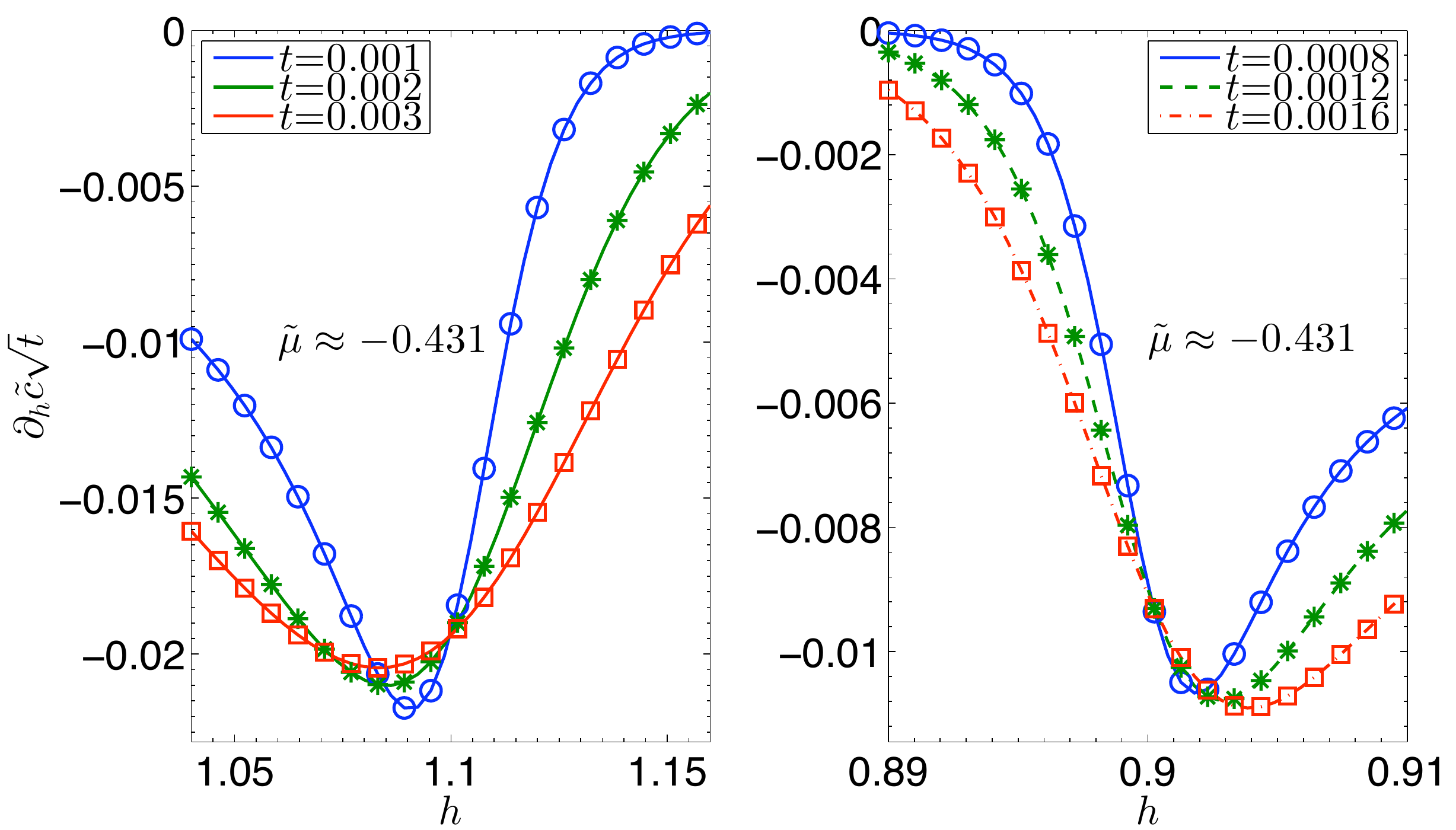}
	  \renewcommand{\figurename}{Supplementary Figure}     
      \caption{Scaling behaviour of the derivative of  contact $\partial_h
  \tilde{c} \sqrt{t}$ vs external field $h$. The left (right)  panel shows the 
   intersection of the derivatives of contact  at different temperatures near the phase transition  F-PP (P-PP).  Here the critical field $h_\textrm{c} = 1.1$ and $h_\textrm{c} = 0.9$, respectively. This plot read off the critical dynamics exponent $z=2$ and correlation length exponent $\nu=1/2$ 
  respectively.}\label{Supplementary-fig:c-h}
\end{figure}

{F-PP:} From the phase F to the phase PP, the critical point is $h_\textrm{c}
= 1 + 2 \left( \tilde{\mu} + h / 2 \right)  \left( 1 - \frac{2 \sqrt{2}}{3
\pi}  \left( \tilde{\mu} + h / 2 \right)^{\frac{1}{2}} \right)$. Near the
critical point we have 
\begin{equation} 
  \tilde{A}_\textrm{u} \approx \left( h - h_\textrm{c} \right) / 2 + \alpha_1, \,
  \tilde{A}_\textrm{b} \approx \beta \left( h - h_\textrm{c} \right),\label{C-h-Scaling-F-PP}
  \end{equation} 
where $\alpha_1 = \left[ \frac{3 \sqrt{2}}{4} \pi \left( \tilde{\mu} + 1 / 2
\right) \right]^{\frac{2}{3}} - \frac{16}{3 \sqrt{2} \pi}  \left( \tilde{\mu}
+ 1 / 2 \right)^{\frac{3}{2}}$ and $\beta = \frac{1}{\sqrt{2} \pi}  \left[ 3
\sqrt{2} \pi \left( 2 \tilde{\mu} + 1 \right) \right]^{\frac{1}{3}}$. 
With the help of these function,  we obtain
\begin{equation}
  \partial_h \widetilde{c} = t^{- \frac{1}{2}} \tmop{Li}_{-
  \frac{1}{2}} \left( - e^{\frac{\beta \left( h - h_\textrm{c} \right)}{t}} \right) 
  \left( \frac{\sqrt{2}}{\pi^{3 / 2}} \alpha_1^{1 / 2} - \frac{2}{\pi^{5 / 2}}
  \alpha_1 - \frac{2 \sqrt{2}}{3 \pi^{3 / 2}} \alpha_1^{3 / 2} \right)
\end{equation}
{P-PP:} From the phase $P$ to the phase $PP$, the critical point is $h_\textrm{c}
= - 2 \tilde{\mu} + \frac{16 \sqrt{2}}{3 \pi}  \left( \tilde{\mu} + 1 / 2
\right)^{\frac{3}{2}}$. Near the critical point we have $
  \tilde{A}_\textrm{u} \approx \left( h - h_\textrm{c} \right) / 2, \,  \tilde{A}_\textrm{b}
  \approx \alpha $
  where $\alpha = 2 \tilde{\mu} + 1 - \frac{2}{3 \pi}  \left( 2 \tilde{\mu} + 1
\right)^{\frac{3}{2}}$.
We obtain
\begin{equation}
  \partial_h \widetilde{c} = t^{- \frac{1}{2}} \tmop{Li}_{-
  \frac{1}{2}} \left( - e^{\frac{h - h_\textrm{c}}{2 t}} \right) \left(
  \frac{1}{\sqrt{2} \pi^{\frac{3}{2}}} \alpha^{\frac{1}{2} } - \frac{5}{2 \sqrt{2} \pi^{\frac{5}{2}}} \alpha - \frac{2 \sqrt{2}}{3 \pi^{\frac{3}{2} }} \alpha^{\frac{3}{2}} +
  \frac{9}{\sqrt{2} \pi^{\frac{7}{2} }} \alpha^{\frac{7}{2}} \right).
\end{equation}
The above result of the scaling function  in term of $h$  can be also cast into the  universal form 

\begin{eqnarray}
   \partial_{\tilde{h} }\widetilde{c} &=&
  \widetilde{c}_{h0} + \lambda_h T^{\left( d / z
  \right) + 1 - \left( 2 / \nu z \right)}  \mathcal{K} \left( \frac{h-
  h_\textrm{c}}{t^{1 / \nu z}} \right).\label{sderivativeh}
\end{eqnarray}
Here $ \widetilde{c}_{h0}$ and $ \lambda_h$ are constant. 

Supplementary Figure 3 shows  divergent behaviours of contact near the phase transitions {P-PP} and {PP-F} driven by the magnetic field $h$, see supplementary equation (14).  Moreover, contact and its derivatives with respect to $h$  at different temperatures must intersect at the critical point. This feature can be used to map out the phase boundaries from the trapped gas at nite temperatures. Supplementary Figure 4 shows the scaling behaviour described by Supplementary equations  (29), (30) and (31).\\

\noindent{\bf \large Supplementary References}

\end{document}